\pdfoutput=1

\documentclass[12pt,a4paper]{article}

\usepackage{ifthen} 
\newboolean{pdflatex}
\setboolean{pdflatex}{False} 

\newboolean{articletitles}
\setboolean{articletitles}{true} 

\newboolean{uprightparticles}
\setboolean{uprightparticles}{false} 

\newboolean{inbibliography}
\setboolean{inbibliography}{false} 

\usepackage{microtype}
\usepackage{lineno}  
\usepackage{xspace} 
\usepackage{caption} 

\usepackage{graphicx}  
\usepackage{color}
\usepackage{colortbl}
\graphicspath{{./figs/}} 

\usepackage{amsmath} 
\usepackage{amssymb}
\usepackage{amsfonts}
\usepackage{upgreek} 

\newcommand*\patchAmsMathEnvironmentForLineno[1]{%
\expandafter\let\csname old#1\expandafter\endcsname\csname #1\endcsname
\expandafter\let\csname oldend#1\expandafter\endcsname\csname
end#1\endcsname
 \renewenvironment{#1}%
   {\linenomath\csname old#1\endcsname}%
   {\csname oldend#1\endcsname\endlinenomath}%
}
\newcommand*\patchBothAmsMathEnvironmentsForLineno[1]{%
  \patchAmsMathEnvironmentForLineno{#1}%
  \patchAmsMathEnvironmentForLineno{#1*}%
}
\AtBeginDocument{%
\patchBothAmsMathEnvironmentsForLineno{equation}%
\patchBothAmsMathEnvironmentsForLineno{align}%
\patchBothAmsMathEnvironmentsForLineno{flalign}%
\patchBothAmsMathEnvironmentsForLineno{alignat}%
\patchBothAmsMathEnvironmentsForLineno{gather}%
\patchBothAmsMathEnvironmentsForLineno{multline}%
\patchBothAmsMathEnvironmentsForLineno{eqnarray}%
}

\usepackage{hyperref}    
\usepackage[all]{hypcap} 

\def\lhcb {\mbox{LHCb}\xspace}

\def\babar  {\mbox{BaBar}\xspace}

\ifthenelse{\boolean{uprightparticles}}%
{

 \def\Pmu         {\ensuremath{\upmu}\xspace}

 \def\Ppi         {\ensuremath{\uppi}\xspace}

 \def\PDelta      {\ensuremath{\Delta}\xspace}                 
 \def\PXi      {\ensuremath{\Xi}\xspace}                 
 \def\PLambda      {\ensuremath{\Lambda}\xspace}                 
 \def\PSigma      {\ensuremath{\Sigma}\xspace}                 
 \def\POmega      {\ensuremath{\Omega}\xspace}                 
 \def\PUpsilon      {\ensuremath{\Upsilon}\xspace}                 
 
 \def\PB      {\ensuremath{\mathrm{B}}\xspace}                 
                  
 \def\PD      {\ensuremath{\mathrm{D}}\xspace}

 \def\PK      {\ensuremath{\mathrm{K}}\xspace}

 \def\Pi      {\ensuremath{\mathrm{i}}\xspace}

 \def\Pp      {\ensuremath{\mathrm{p}}\xspace}

}
{

 \def\Pmu         {\ensuremath{\mu}\xspace}

 \def\Ppi         {\ensuremath{\pi}\xspace}

 \mathchardef\PDelta="7101
 \mathchardef\PXi="7104
 \mathchardef\PLambda="7103
 \mathchardef\PSigma="7106
 \mathchardef\POmega="710A
 \mathchardef\PUpsilon="7107
                  
 \def\PB      {\ensuremath{B}\xspace}                 
                  
 \def\PD      {\ensuremath{D}\xspace}

 \def\PK      {\ensuremath{K}\xspace}

 \def\Pi      {\ensuremath{i}\xspace}

 \def\Pp      {\ensuremath{p}\xspace}

}

\makeatletter
\ifcase \@ptsize \relax
  \newcommand{\miniscule}{\@setfontsize\miniscule{4}{5}}
\or
  \newcommand{\miniscule}{\@setfontsize\miniscule{5}{6}}
\or
  \newcommand{\miniscule}{\@setfontsize\miniscule{5}{6}}
\fi
\makeatother

\DeclareRobustCommand{\optbar}[1]{\shortstack{{\miniscule (\rule[.5ex]{1.25em}{.18mm})}
  \\ [-.7ex] $#1$}}

\def\mun        {{\ensuremath{\Pmu^-}}\xspace} 

\def\pion   {{\ensuremath{\Ppi}}\xspace}

\def\pip    {{\ensuremath{\pion^+}}\xspace}
\def\pim    {{\ensuremath{\pion^-}}\xspace}

\def\kaon    {{\ensuremath{\PK}}\xspace}
  \def\Kbar    {{\kern 0.2em\overline{\kern -0.2em \PK}{}}\xspace}

\def\KorKbar    {\kern 0.18em\optbar{\kern -0.18em K}{}\xspace}

\def\Kp      {{\ensuremath{\kaon^+}}\xspace}
\def\Km      {{\ensuremath{\kaon^-}}\xspace}

\def\KS      {{\ensuremath{\kaon^0_{\rm\scriptscriptstyle S}}}\xspace}

  \def\Dbar    {{\kern 0.2em\overline{\kern -0.2em \PD}{}}\xspace}
\def\D       {{\ensuremath{\PD}}\xspace}

\def\DorDbar    {\kern 0.18em\optbar{\kern -0.18em D}{}\xspace}
\def\Dz      {{\ensuremath{\D^0}}\xspace}
\def\Dzb     {{\ensuremath{\Dbar{}^0}}\xspace}

\def\Dstarp  {{\ensuremath{\D^{*+}}}\xspace}

\def\B       {{\ensuremath{\PB}}\xspace}
\def\Bbar    {{\ensuremath{\kern 0.18em\overline{\kern -0.18em \PB}{}}}\xspace}

\def\BorBbar    {\kern 0.18em\optbar{\kern -0.18em B}{}\xspace}

  \def\Y#1S{\ensuremath{\PUpsilon{(#1S)}}\xspace}

\def\proton      {{\ensuremath{\Pp}}\xspace}

\def\Lbar        {{\ensuremath{\kern 0.1em\overline{\kern -0.1em\PLambda}}}\xspace}
\def\LorLbar    {\kern 0.18em\optbar{\kern -0.18em \PLambda}{}\xspace}

\def\to                 {\ensuremath{\rightarrow}\xspace}

\def\CP                {{\ensuremath{C\!P}}\xspace}

\def\AT#1     {\ensuremath{A_{\mathrm{T}}^{#1}}\xspace}           

\def\C#1      {\ensuremath{\mathcal{C}_{#1}}\xspace}                       
\def\Cp#1     {\ensuremath{\mathcal{C}_{#1}^{'}}\xspace}                    
\def\Ceff#1   {\ensuremath{\mathcal{C}_{#1}^{\mathrm{(eff)}}}\xspace}        
\def\Cpeff#1  {\ensuremath{\mathcal{C}_{#1}^{'\mathrm{(eff)}}}\xspace}       
\def\Ope#1    {\ensuremath{\mathcal{O}_{#1}}\xspace}                       
\def\Opep#1   {\ensuremath{\mathcal{O}_{#1}^{'}}\xspace}                    

\newcommand{\tev}{\ifthenelse{\boolean{inbibliography}}{\ensuremath{~T\kern -0.05em eV}\xspace}{\ensuremath{\mathrm{\,Te\kern -0.1em V}}}\xspace}
\newcommand{\gev}{\ensuremath{\mathrm{\,Ge\kern -0.1em V}}\xspace}
\newcommand{\mev}{\ensuremath{\mathrm{\,Me\kern -0.1em V}}\xspace}
\newcommand{\kev}{\ensuremath{\mathrm{\,ke\kern -0.1em V}}\xspace}
\newcommand{\ev}{\ensuremath{\mathrm{\,e\kern -0.1em V}}\xspace}
\newcommand{\gevc}{\ensuremath{{\mathrm{\,Ge\kern -0.1em V\!/}c}}\xspace}
\newcommand{\mevc}{\ensuremath{{\mathrm{\,Me\kern -0.1em V\!/}c}}\xspace}
\newcommand{\gevcc}{\ensuremath{{\mathrm{\,Ge\kern -0.1em V\!/}c^2}}\xspace}
\newcommand{\gevgevcccc}{\ensuremath{{\mathrm{\,Ge\kern -0.1em V^2\!/}c^4}}\xspace}
\newcommand{\mevcc}{\ensuremath{{\mathrm{\,Me\kern -0.1em V\!/}c^2}}\xspace}

\def\mum  {\ensuremath{{\,\upmu\rm m}}\xspace}

\def\invfb   {\ensuremath{\mbox{\,fb}^{-1}}\xspace}

\def\ps   {\ensuremath{{\rm \,ps}}\xspace}

\newcommand{\stat}{\ensuremath{\mathrm{\,(stat)}}\xspace}
\newcommand{\syst}{\ensuremath{\mathrm{\,(syst)}}\xspace}

\newcommand{\chisq}{\ensuremath{\chi^2}\xspace}
\newcommand{\chisqndf}{\ensuremath{\chi^2/\mathrm{ndf}}\xspace}

\def\gsim{{~\raise.15em\hbox{$>$}\kern-.85em
          \lower.35em\hbox{$\sim$}~}\xspace}
\def\lsim{{~\raise.15em\hbox{$<$}\kern-.85em
          \lower.35em\hbox{$\sim$}~}\xspace}

\def\pt         {\mbox{$p_{\rm T}$}\xspace}

\def\evtgen     {\mbox{\textsc{EvtGen}}\xspace}

\def\geant      {\mbox{\textsc{Geant4}}\xspace}

\def\pythia     {\mbox{\textsc{Pythia}}\xspace}

\def\tell1  {TELL1\xspace}
\def\ukl1   {UKL1\xspace}

\newcommand{\ie}{\mbox{\itshape i.e.}\xspace}

\def\atv	{\ensuremath{a_{\CP}^{\text{\T-odd}}}\xspace} 
\def\at	{\ensuremath{A_{T}}\xspace} 
\def\atb	{\ensuremath{\overline{A}_{T}}\xspace} 
\def\ct	{\ensuremath{C_{T}}\xspace} 
\def\ctb	{\ensuremath{\overline{C}_{T}}\xspace}

\def\DzToKKpipi	{\ensuremath{\Dz\to\Kp\Km\pip\pim}\xspace}
\def\DzToKpipipi{\ensuremath{\Dz\to\Km\pip\pim\pip}\xspace}
\def\DzToKsKK	{\ensuremath{\Dz\to\KS\Kp\Km}\xspace}

\def\BtoDmuX    {\ensuremath{\B\to\Dz\mun X}\xspace}

\def\CPV                {{\ensuremath{C\!PV}}\xspace}
\def\T	        {\ensuremath{T}\xspace}

\def\beq {\begin{equation}}
\def\eeq {\end{equation}}

\usepackage{cite} 
\usepackage{mciteplus}

\usepackage{longtable} 

\usepackage{booktabs}

\begin{document}

\renewcommand{\thefootnote}{\fnsymbol{footnote}}
\setcounter{footnote}{1}

\begin{titlepage}
\pagenumbering{roman}

\vspace*{-1.5cm}
\centerline{\large EUROPEAN ORGANIZATION FOR NUCLEAR RESEARCH (CERN)}
\vspace*{1.5cm}
\hspace*{-0.5cm}
\begin{tabular*}{\linewidth}{lc@{\extracolsep{\fill}}r}
\ifthenelse{\boolean{pdflatex}}
{\vspace*{-2.7cm}\mbox{\!\!\!\includegraphics[width=.14\textwidth]{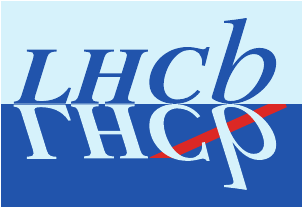}} & &}%
{\vspace*{-1.2cm}\mbox{\!\!\!\includegraphics[width=.12\textwidth]{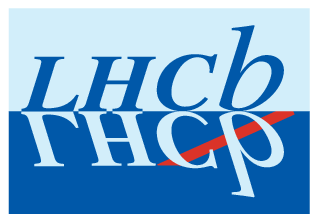}} & &}%
\\
 & & CERN-PH-EP-2014-194 \\  
 & & LHCb-PAPER-2014-046 \\  
 & & October 1, 2014 \\
 & & \\
\end{tabular*}

\vspace*{3.9cm}

{\bf\boldmath\huge
\begin{center}
  Search for \CP violation using $T$-odd correlations in
  $\Dz\to\Kp\Km\pip\pim$ decays
\end{center}
}

\vspace*{1.7cm}

\begin{center}
The LHCb collaboration\footnote{Authors are listed at the end of this paper.}
\end{center}

\vspace{\fill}

\begin{abstract}
  \noindent
  A search for \CP violation using \T-odd correlations is performed using the four-body \DzToKKpipi decay, selected from semileptonic \B decays.
  The data sample corresponds to integrated luminosities of 1.0\invfb and 2.0\invfb recorded at the centre-of-mass energies of 7\tev and 8\tev, respectively.
  The \CP-violating asymmetry \atv is measured to be $(0.18\pm 0.29\stat\pm 0.04\syst)\%$.
 Searches for \CP violation in different regions of phase space of the four-body decay, and as a function of the \Dz decay time, are also presented.
  No significant deviation from the \CP conservation hypothesis is found.
\end{abstract}

\vspace*{2.0cm}

\begin{center}
  Published in JHEP
\end{center}

\vspace{\fill}

{\footnotesize 
\centerline{\copyright~CERN on behalf of the \lhcb collaboration, license \href{http://creativecommons.org/licenses/by/4.0/}{CC-BY-4.0}.}}
\vspace*{2mm}

\end{titlepage}


\newpage
\setcounter{page}{2}
\mbox{~}
\newpage

\cleardoublepage



\renewcommand{\thefootnote}{\arabic{footnote}}
\setcounter{footnote}{0}


\pagestyle{plain} 
\setcounter{page}{1}
\pagenumbering{arabic}

\section{Introduction}
\label{intro}
Violation of the \CP symmetry in charm decays is expected to be very small in the Standard Model (SM)~\cite{Bianco:2003vb, PhysRevD.75.036008}, however, asymmetries at a few times 10$^{-3}$ within the SM cannot be excluded according to recent calculations~\cite{Feldmann:2012js, Brod:2011re, Bhattacharya:2012ah}. 
A significant excess of \CP violation (\CPV) with respect to the theoretical predictions would be a signature of physics beyond the SM.
The study of \CPV in singly Cabibbo-suppressed charm decays is uniquely sensitive to physics beyond the SM, in particular through new contributions in strong penguin and chromomagnetic dipole operators~\cite{PhysRevD.75.036008}.
The analysis of singly Cabibbo-suppressed \DzToKKpipi\footnote{
Throughout this paper the use of charge conjugate reactions is implied, unless otherwise indicated.} decays allows localised \CPV in different regions 
 of phase space to be probed. This approach enhances the sensitivity due to several interfering amplitudes with 
 different relative strong phases contributing to the decay.

The analysis in Ref.~\cite{tagkey2013623} quotes a $p$-value of 9.1\%  
for the compatibility with the \CP conservation hypothesis, using $D^{*}$-tagged 
promptly-produced \Dz mesons. In the present analysis, 
a sample of \DzToKKpipi decays, selected from semileptonic \B decays, is used 
to measure a \CP-violating parameter based on $T$-odd correlations characterised by different sensitivity to \CPV~\cite{Valencia:1988it,Datta:2003mj}.
Using triple products of final state particle momenta in the \Dz centre-of-mass frame,  
$\ct \equiv \vec{p}_{\Kp} \cdot (\vec{p}_{\pip}\times\vec{p}_{\pim})$ for 
 \Dz and $\ctb \equiv \vec{p}_{\Km} \cdot (\vec{p}_{\pim}\times\vec{p}_{\pip})$ for \Dzb decays, 
 two \T-odd observables,
\beq
\label{eq:AT_ATb}
\at \equiv \frac{\Gamma_{\Dz}(\ct>0) - \Gamma_{\Dz}(\ct<0)}{\Gamma_{\Dz}(\ct>0) + \Gamma_{\Dz}(\ct<0)}, \quad \quad
\atb \equiv \frac{\Gamma_{\Dzb}(-\ctb>0) - \Gamma_{\Dzb}(-\ctb<0)}{\Gamma_{\Dzb}(-\ctb>0) + \Gamma_{\Dzb}(-\ctb<0)},
\eeq
can be studied~\cite{Bigi:2001sg}, where $\Gamma_{\Dz}$ ($\Gamma_{\Dzb}$) is the partial decay width of \Dz (\Dzb) decays to $\Kp\Km\pip\pim$ in the indicated \ct (\ctb) range.
However, because final state interaction (FSI) effects could introduce fake 
asymmetries~\cite{Bigi:2001sg,PhysRevD.84.096013}, 
these are not theoretically clean \CP-violating observables. 
A well defined \CP-violating observable is  
\beq
\label{eq:ATv}
\atv \equiv \frac{1}{2}(\at-\atb),
\eeq
as FSI effects cancel out in the difference.
In contrast to the asymmetry between the phase-space integrated rates
in a $\Dz\to V_1 V_2$ decay (where $V_i$ indicates a vector meson),
\atv is sensitive to \CP violation in interference between even- and odd- partial waves of the $V_1V_2$ system\cite{Valencia:1988it}.

Previous measurements of \atv are compatible with no \CPV: FOCUS measured 
$\atv = (1.0 \pm 5.7 \pm 3.7)\%$~\cite{Link:2005th}, and \babar measured 
$\atv = (0.10 \pm 0.51 \pm 0.44)\%$~\cite{delAmoSanchez:2010xj}.
 The physics observables, \at, \atb, and \atv are by construction insensitive to \Dz/\Dzb production asymmetries,
 detector- and  reconstruction-induced charge asymmetries. 
The measurement described in this paper determines the \CP-violating observable \atv with 
an improved precision.
For the first time, \atv is measured in different regions of phase space
 and in bins of \Dz decay time, allowing to probe for \CP violation both in the decay amplitude and in its interference with the mixing amplitude.

\section{Detector}
\label{sec:detector}

 The LHCb detector~\cite{Alves:2008zz} is a single-arm forward spectrometer covering the pseudorapidity
 range $2<\eta<5$, designed for the study of particles containing $b$ or $c$ quarks. The detector
 includes a high-precision tracking system consisting of a silicon-strip vertex detector
 surrounding the \proton\proton interaction region~\cite{LHCbVELOGroup:2014uea}, 
a large-area silicon-strip detector located
 upstream of a dipole magnet with a bending power of about $4{\rm\,Tm}$, and three stations of
 silicon-strip detectors and straw drift tubes~\cite{LHCb-DP-2013-003} placed downstream of the magnet. The
 tracking system provides a measurement of momentum, $p$, with a relative uncertainty
 that varies from 0.4\% at low momentum to 0.6\% at $100\gevc$. The minimum distance
 of a track to a primary vertex, the impact parameter, is measured with a resolution
 of ($15 + (29\gevc)/\pt$)\mum, where \pt is the component of $p$ transverse to the beam.
 Different types of charged hadrons are distinguished using information from
 two ring-imaging Cherenkov (RICH) detectors~\cite{LHCb-DP-2012-003}. Photon, electron and hadron candidates are
 identified by a calorimeter system consisting of scintillating-pad and preshower detectors,
 an electromagnetic calorimeter and a hadronic calorimeter. Muons are identified by a
 system composed of alternating layers of iron and multiwire proportional chambers~\cite{LHCb-DP-2012-002}.
 The trigger~\cite{LHCb-DP-2012-004} consists of a hardware stage, based on information from the calorimeter
 and muon systems, followed by a software stage, which applies a full event reconstruction.

Events are required to pass both hardware and software trigger selections.
The software trigger identifies  $\DzToKKpipi$ (signal) and $\DzToKpipipi$ (control sample) 
events from $\B \to \Dz \mun X$
 decays, where $X$ indicates any system composed of charged and neutral particles, by requiring a four-track secondary vertex with a scalar sum of \pt of the tracks greater than $1.8\gevc$. 
The \Dz daughter tracks are required to have $\pt>0.3\gevc$ and momentum $p>2\gevc$. 
The muon track is selected with $\pt>1.2\gevc$ and $p>2\gevc$. 
Tracks have to be compatible with detached decay vertices of $B$ and $\Dz$ decays. 
Therefore, a requirement is imposed for all the tracks in the signal candidate on the $\chisq_{\text{IP}}$, \ie the difference in $\chi^2$ of a given primary vertex reconstructed with and without the considered particle, to be greater than 9. 
The invariant mass of the $\Dz\mu$ system is required to be less than $6.2\gevcc$. 

In the simulation, \proton\proton collisions are generated using \pythia~\cite{Sjostrand:2006za,Sjostrand:2007gs} 
with a specific LHCb configuration~\cite{LHCb-PROC-2010-056}. 
Decays of hadronic particles are described by \evtgen~\cite{Lange:2001uf}.
The interaction of the generated particles with the detector and its response are implemented using the \geant toolkit~\cite{Allison:2006ve,Agostinelli:2002hh} 
as described in Ref.~\cite{LHCb-PROC-2011-006}.

\section{Selection}
\label{sec:selection}

The analysis is based on data recorded by the \lhcb experiment, at center-of-mass 
energies of $7\tev$ and $8\tev$, corresponding to integrated luminosities 
of $1.0\invfb$ and $2.0\invfb$, respectively. 

The \Dz candidates are formed from combinations of kaon and pion candidate tracks and then 
 combined with a muon candidate track to reconstruct the semileptonic \B decay.
The flavour of the \Dz is identified by the charge of the muon, \ie a negative charge
identifies a \Dz meson  and a positive charge identifies a \Dzb meson.
The information from the RICH system is used to distinguish between kaons and pions, while the
information from the muon system is used to identify muon candidates. 
The \Dz meson decay vertex is required to be downstream of the \B decay vertex.
The invariant mass of the $\Dz \mu$ system is required to be in the range
 $[2.6, 5.2] \gevcc$.  

Two main sources of peaking background in $m(\Kp\Km\pip\pim)$, the reconstructed invariant mass
 of \Dz candidates, are present and consist of
 \DzToKsKK decays, and \DzToKKpipi decays from \Dz mesons that originate at the interaction point, 
referred to as ``prompt'' charm decays in the following.
The small component of \DzToKsKK events is vetoed by requiring  
 the invariant mass of the \pip\pim system to be more than $3\sigma$ away from 
the  known \KS mass~\cite{PDG2012}, where $\sigma=4.5\mevcc$ is the resolution determined 
from the fit to data.
The contribution of prompt charm decays is estimated by fitting 
the distribution of the logarithm of the $\chisq_{\textrm{IP}}$ of the \Dz meson.
The prompt component and the signal component from semileptonic \B decays accumulate at 0 and at 5, respectively.
The fraction of prompt \Dz decays in our sample is measured to be $f_{\textrm{prompt}}=(1.20\pm0.08)\%$ and the effect of its presence is accounted for as a systematic uncertainty.
The distributions of \Dz decays where pions and kaons have not been correctly identified have been studied and do not peak
 in $m(\Kp\Km\pip\pim)$.
When multiple candidates are reconstructed, one candidate per event is retained, 
by random choice. This happens in 0.7\% of the events.
The signal yield for \DzToKKpipi from \BtoDmuX decays, 
obtained from an extended maximum likelihood fit to the $m(\Kp\Km\pip\pim)$ 
distribution, is $(171.3\pm0.6)\times10^3$ events with a sample purity of about 75\%.

By using identical kinematic selection criteria as for the signal,
 Cabibbo-favoured \DzToKpipipi decays are also reconstructed with a signal yield of about
of $6.211\times 10^6$ and a purity of about 95\%.
These decays are used for control checks and for assessing systematic uncertainties.

\section{Asymmetry measurements}
\label{sec:analysis}

\begin{figure}[tb]
\centering
\small
\includegraphics[width=0.4\textwidth]{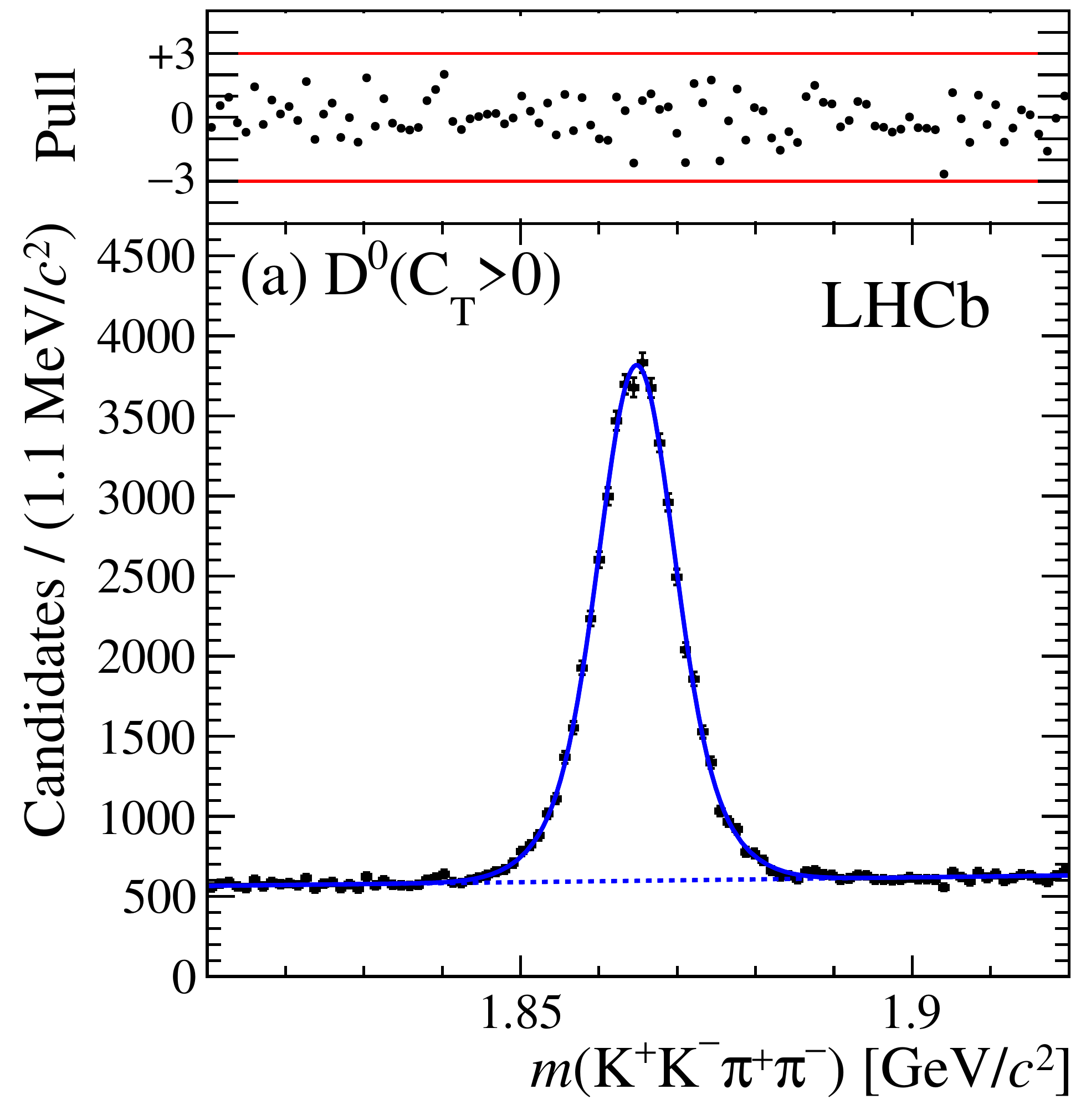}
\includegraphics[width=0.4\textwidth]{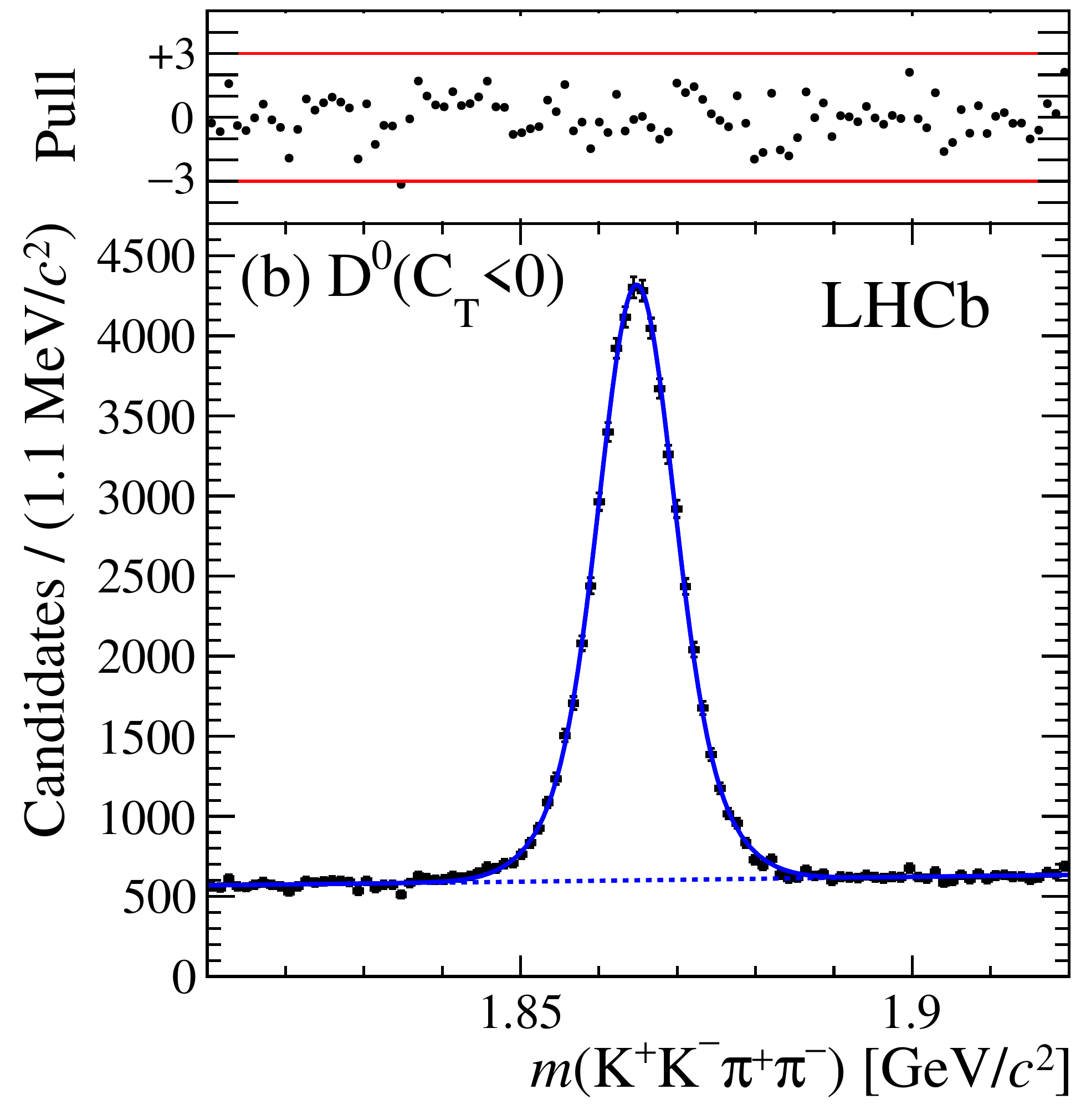}
\includegraphics[width=0.4\textwidth]{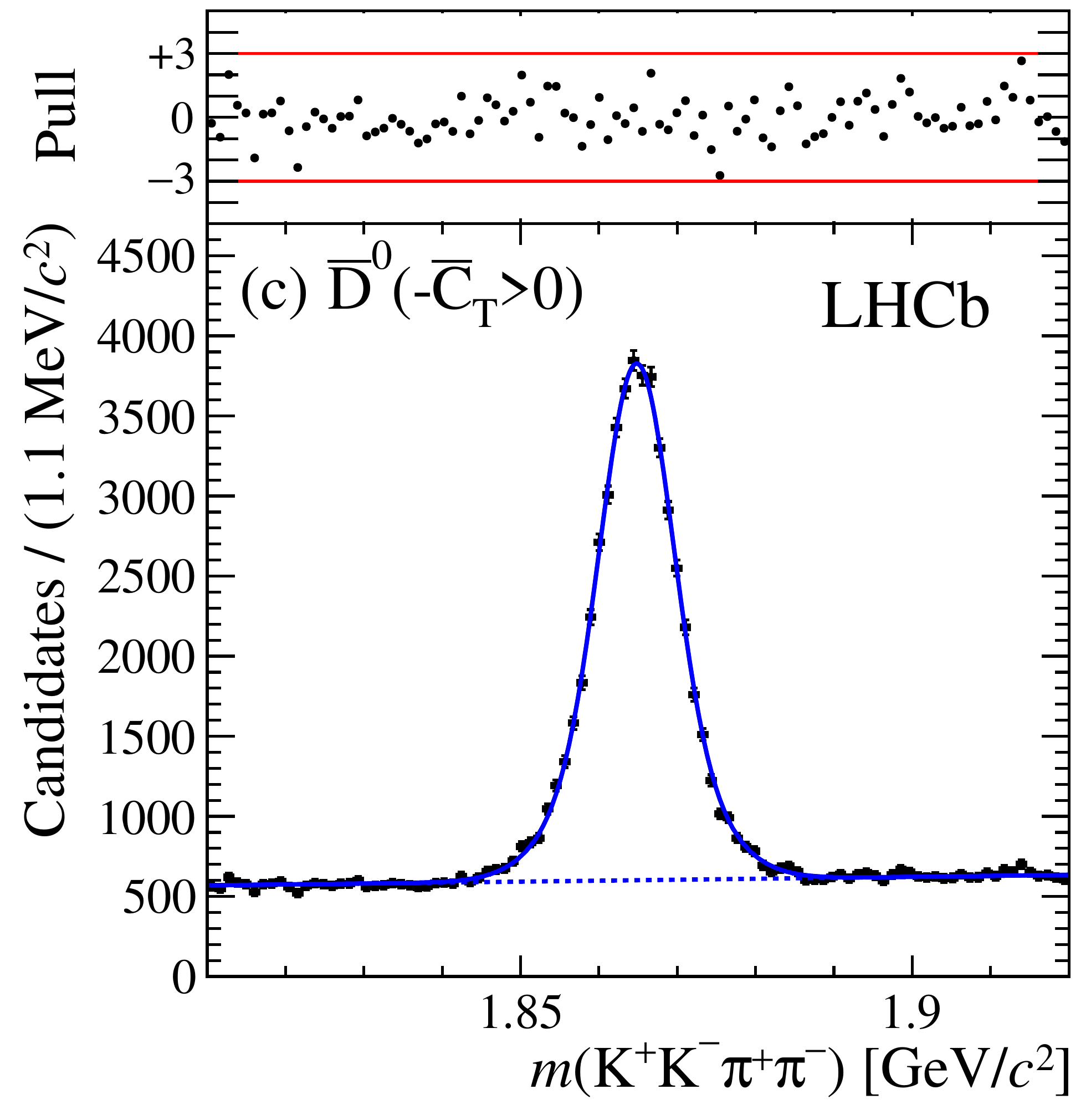}
\includegraphics[width=0.4\textwidth]{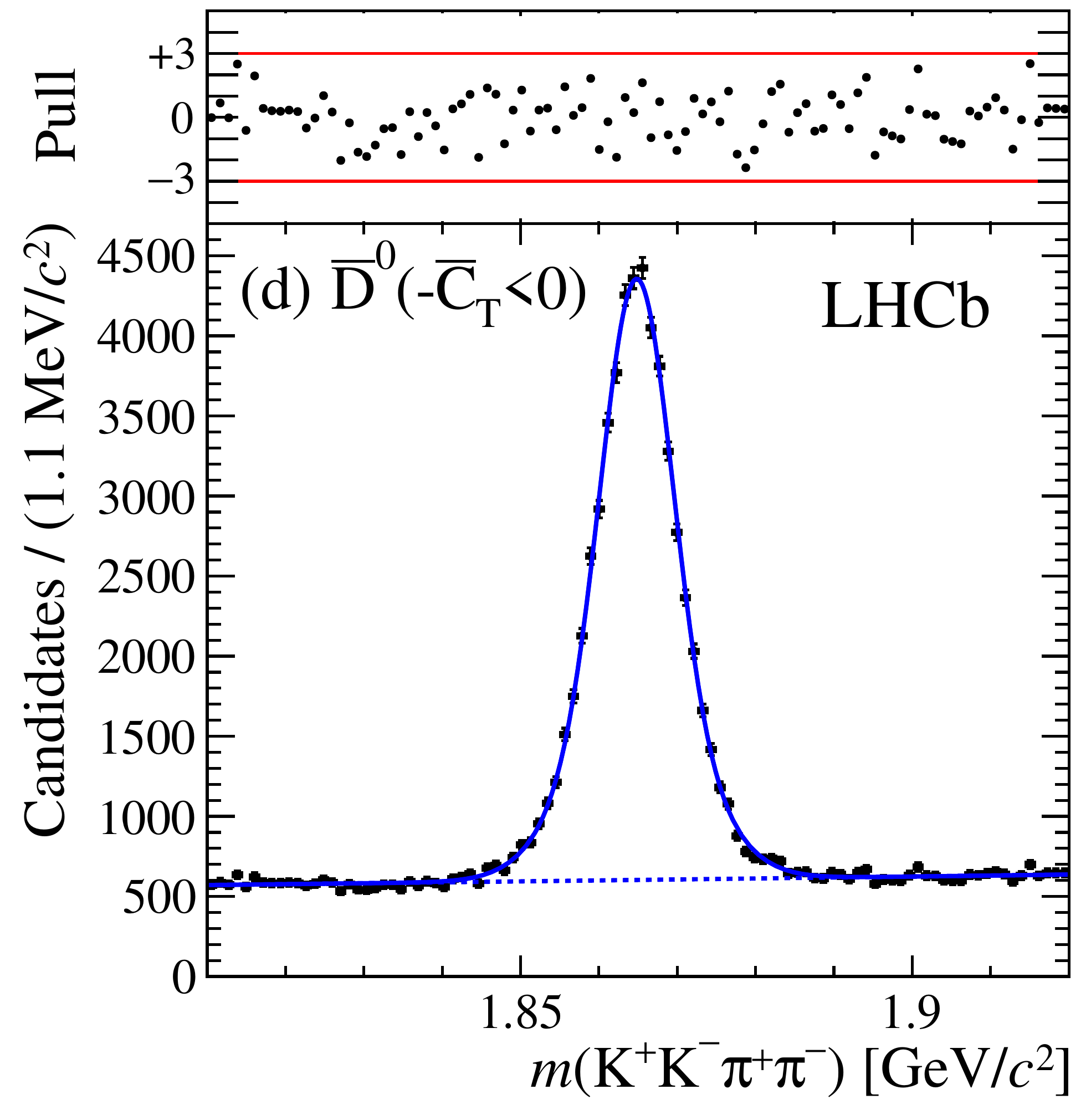}
\caption{\small\label{fig:integrated}Distributions of the $\Kp\Km\pip\pim$ invariant mass in the four samples 
defined by \Dz (\Dzb) flavour and the sign of \ct (\ctb).
The results of the fit 
are overlaid as a solid line, and a dashed line is used for representing the background.
The normalised residuals (pulls) of the difference between the fit results 
and the data points, divided by their uncertainties, are shown on top of each distribution.}
\end{figure}

The selected data sample is split into four subsamples
according to the charge of the muon candidate, which determines the flavour 
 of the \Dz, and the sign of \ct (\ctb).
 The reconstruction efficiencies are equal, within their uncertainties, 
for $\ct>0$ ($-\ctb>0$) and for $\ct<0$ ($-\ctb<0$) according to studies based on simulated events 
and on the $\DzToKpipipi$ control sample.
A simultaneous maximum likelihood fit to the $m(\Kp\Km\pip\pim)$ distribution 
of the four subsamples is used to determine the number 
of signal and background events, and the asymmetries \at and \atb.
The fit model consists of two Gaussian functions with common mean for the signal
 and an exponential function for the background.
The two asymmetries \at and \atb are included in the fit model as 
\begin{equation}
\begin{aligned}
N_{\Dz,\ct>0}&= \frac{1}{2}N_{\Dz}(1+\at),\\
N_{\Dz,\ct<0}&= \frac{1}{2}N_{\Dz}(1-\at),\\
N_{\Dzb,-\ctb>0}&= \frac{1}{2}N_{\Dzb}(1+\atb),\\
N_{\Dzb,-\ctb<0}&= \frac{1}{2}N_{\Dzb}(1-\atb).
\end{aligned}
\end{equation}
The \CP-violating asymmetry \atv is then calculated from \at and \atb.
Negligible correlation is found between these two asymmetries.
The results of the fit are shown in Fig.~\ref{fig:integrated}.
The number of signal decays for each subsample are listed in Table~\ref{tab:nevs}.
\begin{table}
\small
\centering
\caption{\small\label{tab:nevs}Number of signal decays obtained from the fit to data for each of the four samples defined by the \Dz/\Dzb flavour and the sign of \ct or \ctb.}
\begin{tabular}{lc}
\toprule
Sample        	& Signal Decays\\
\midrule
\Dz, $\phantom{-}\ct>0$   	& $39\,628 \pm 256$\\
\Dz, $\phantom{-}\ct<0$   	& $45\,762 \pm 272$\\
\Dzb, $-\ctb>0$  & $39\,709 \pm 256$\\
\Dzb, $-\ctb<0$  & $46\,162 \pm 274$\\
\bottomrule
\end{tabular}

\end{table}

Three different approaches have been followed to search for \CPV:
 a  measurement integrated over the phase space,  measurements in different regions of phase space, 
and measurements as a function of the \Dz decay time.
The results of the first approach are obtained by fitting the full data sample 
and are $\at=(-7.18\pm0.41)\%$, and $\atb=(-7.55\pm0.41)\%$, where the uncertainties are 
statistical only.
The \CP-violating asymmetry calculated from the two partial asymmetries is $\atv=(0.18\pm0.29)\%$.

The relatively large asymmetries observed in \at and \atb are due to FSI effects~\cite{Bigi:2001sg, PhysRevD.84.096013}, which are known to be relevant in charm mesons decays.
These effects are difficult to predict, since they involve non-perturbative strong interactions~\cite{Gronau:1999zt}.
However, experimental measurements provide solid anchor points for future calculations.

The measurement in different regions of the phase space is performed by dividing 
the sample using a binning scheme based on the Cabibbo-Maksimowicz~\cite{Cabibbo:1965zzb} 
variables $m_{\Kp\Km}$, $m_{\pip\pim}$, $\cos(\theta_{\Kp})$, $\cos(\theta_{\pip})$, $\Phi$, 
defined as the $\Kp\Km$ and $\pip\pim$ invariant masses, the cosine of the angle of the \Kp (\pip) 
with respect to the opposite direction to the \Dz momentum in the $\Kp\Km$ ($\pip\pim$) rest frame, 
and the angle between the planes described by the two kaons and pions in the \Dz rest frame, respectively.

\begin{figure}[tb]
\centering
\small
\includegraphics[width=0.75\textwidth]{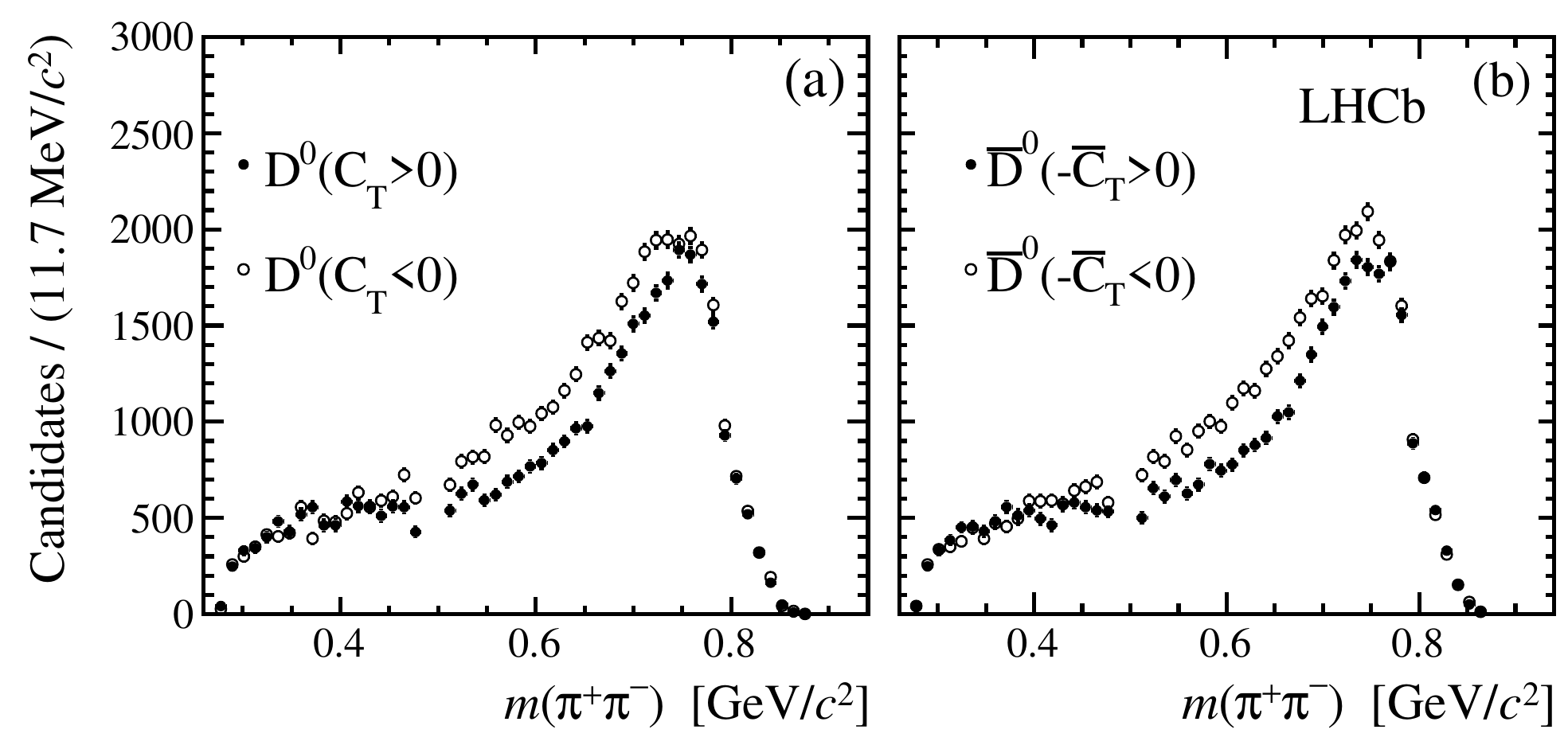}
\includegraphics[width=0.75\textwidth]{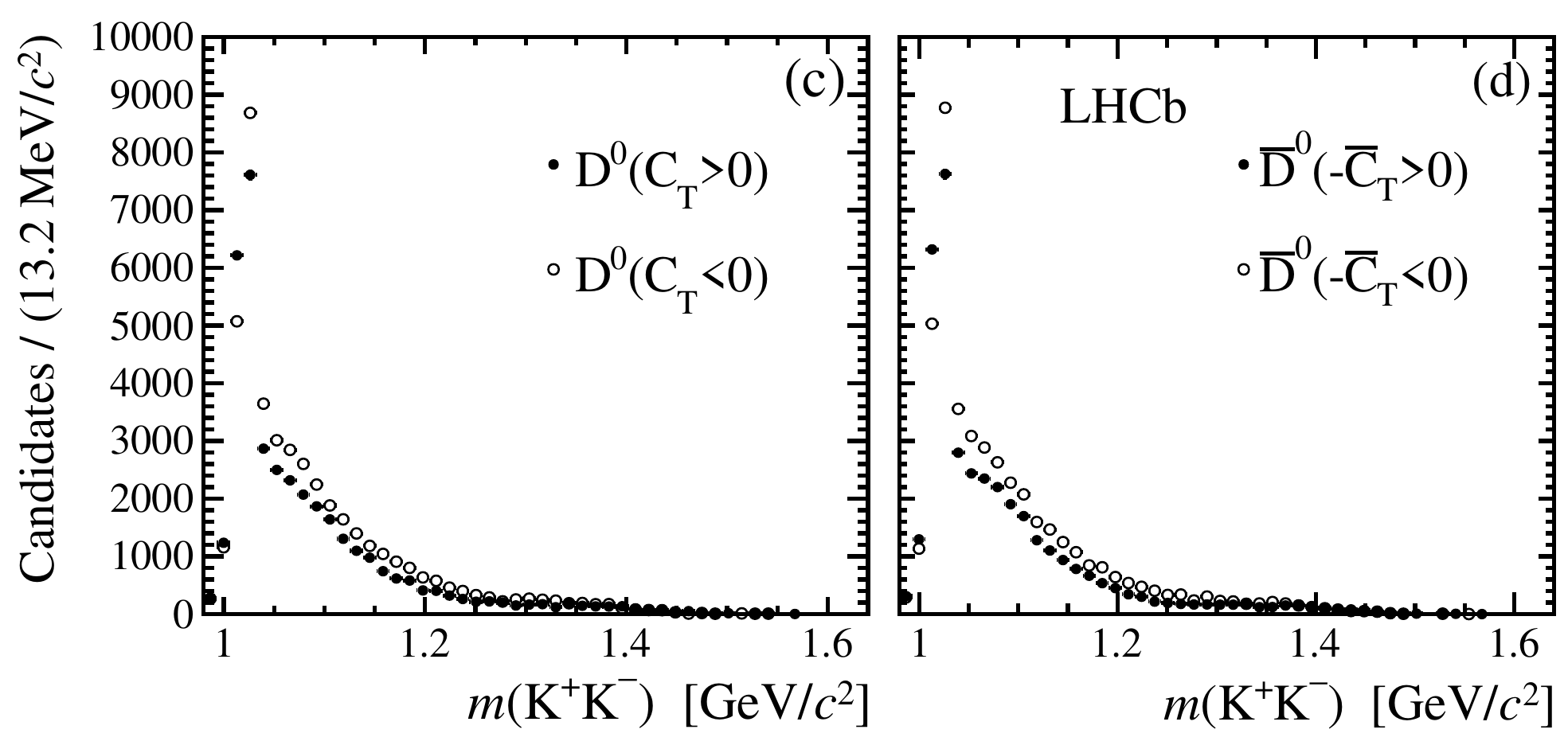}
\caption{\small\label{fig:phspDistributionsI}Sideband-subtracted distributions of \Dz (\Dzb) candidates
in variables of (a, b) $m_{\pip\pim}$  and (c, d) $m_{\Kp\Km}$  for different values of \ct (\ctb). 
The veto for \DzToKsKK decays is visible in the $m_{\pip\pim}$ distribution.
}
\end{figure}

\begin{figure}[tb]
\centering
\small
\includegraphics[width=0.7\textwidth]{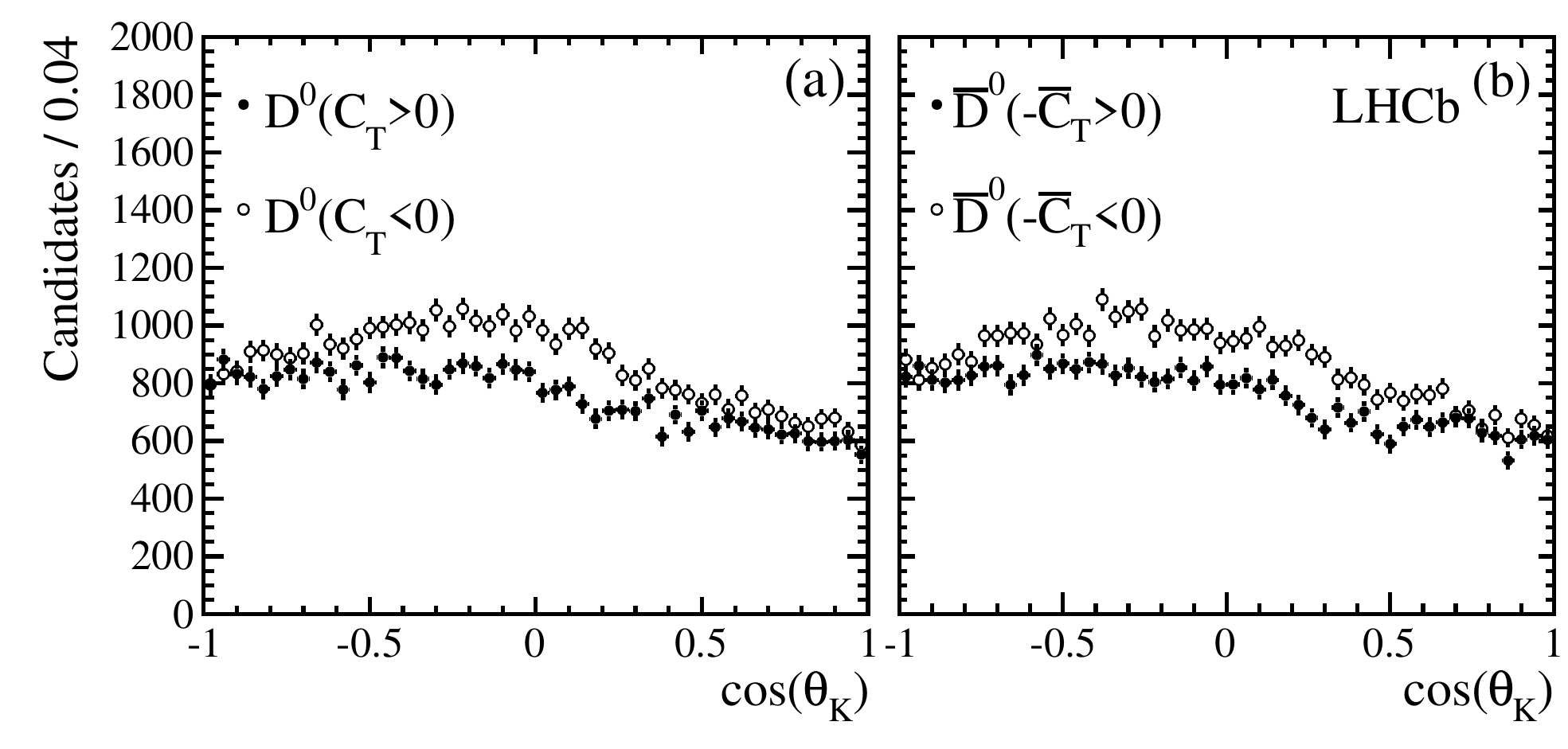}
\includegraphics[width=0.7\textwidth]{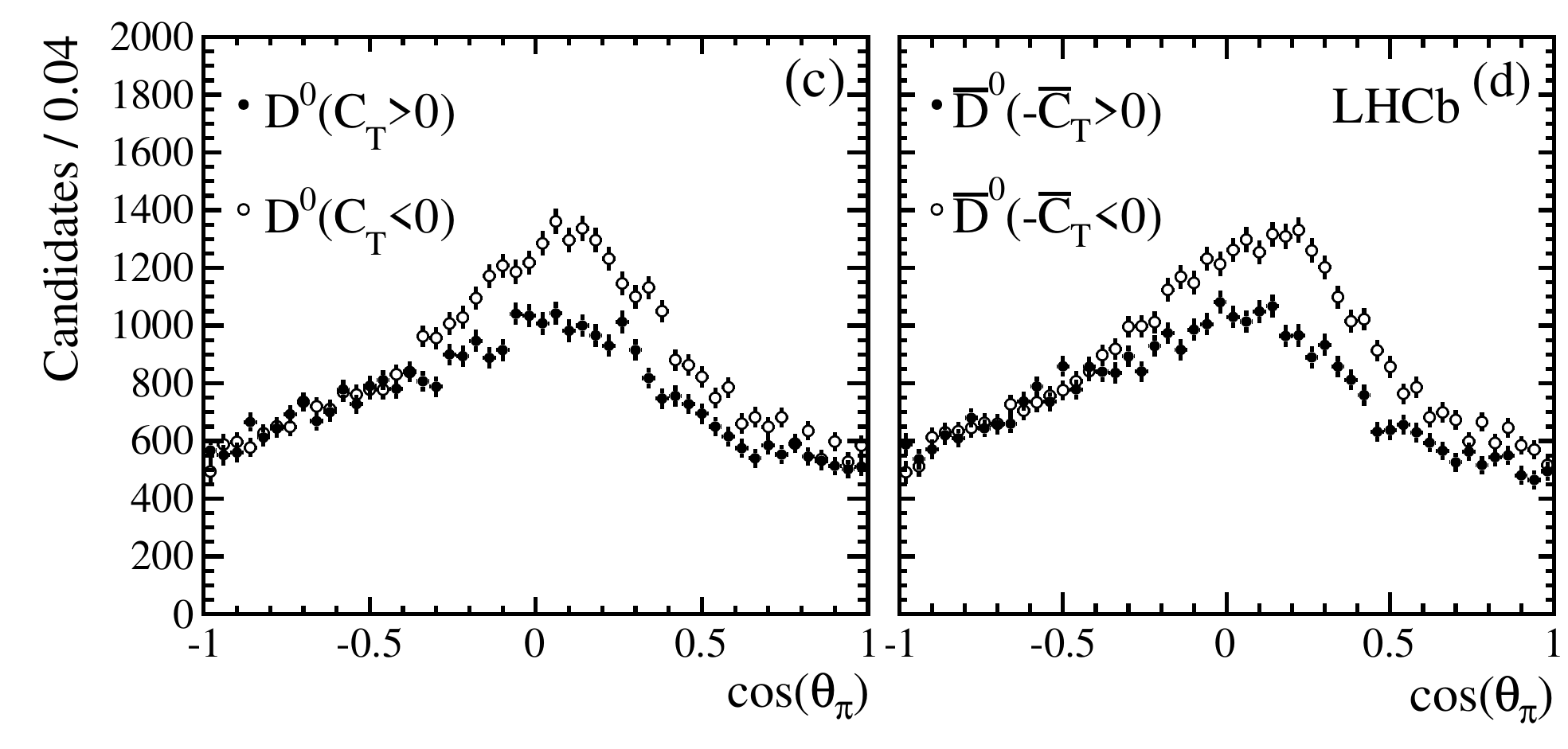}
\includegraphics[width=0.7\textwidth]{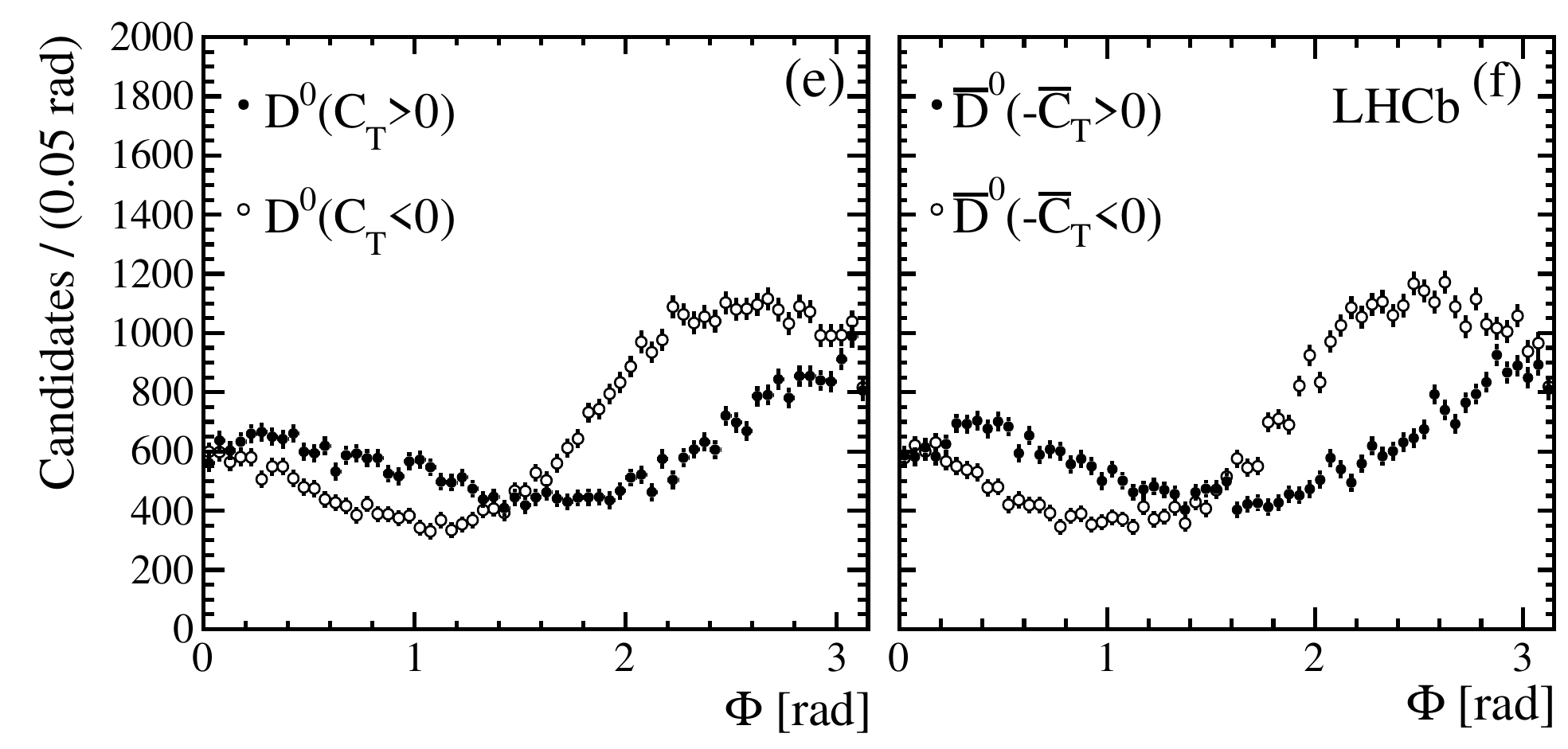}
\caption{\small\label{fig:phspDistributionsII}Sideband-subtracted distributions of \Dz (\Dzb) candidates in variables of (a, b) $\cos(\theta_{\Kp})$,  (c, d) $\cos(\theta_{\pip})$, and (e, f) $\Phi$  for different values of \ct (\ctb). 
The asymmetric distributions with respect to 0 for $\cos(\theta_{\Kp})$ and  $\cos(\theta_{\pip})$ variables, and 
with respect to $\pi/2$ for the $\Phi$ variable, are due to the dynamics of the four-body decay.
}
\end{figure}

The background-subtracted distributions for \Dz (\Dzb) events with $\ct>0$ and $\ct<0$ 
($-\ctb>0$ and $-\ctb<0$) in  $m_{\pip\pim}$ and $m_{\Kp\Km}$ 
are shown in Fig.~\ref{fig:phspDistributionsI}. 
The background subtraction is performed using $m(\Kp\Km\pip\pim)$ sidebands.
Clear indications of $\rho^0 \to \pip\pim$ and $\phi\to\Kp\Km$ resonances are seen in the data.
The distributions of $\cos(\theta_{\Kp})$, $\cos(\theta_{\pip})$, and $\Phi$  variables
are shown in Fig.~\ref{fig:phspDistributionsII}, where FSI-induced differences are clearly evident~\cite{Donoghue:1986nu}.
Effects of \CPV would lead to different distributions for \Dz and \Dzb events.

The phase space is divided in 32 regions such that the number of signal events is similar in each region; 
 the definition of the 32 regions is reported in Table~\ref{tab:32regions} in Appendix~\ref{app:phase_space}. 

The same fit model used for the integrated measurement is separately fitted to data in each bin.
The signal shapes are consistent among different bins, while significant variations are found in the distribution of the combinatorial background.
The distributions of the measured asymmetries in the 32-region binning scheme are shown in Fig.~\ref{fig:phspBinning}
 and the results are reported in Table~\ref{tab:PhaseSpace_Results} in Appendix~\ref{app:phase_space}. 
\begin{figure}[tb]
\centering
\small
\includegraphics[width=0.32\textwidth]{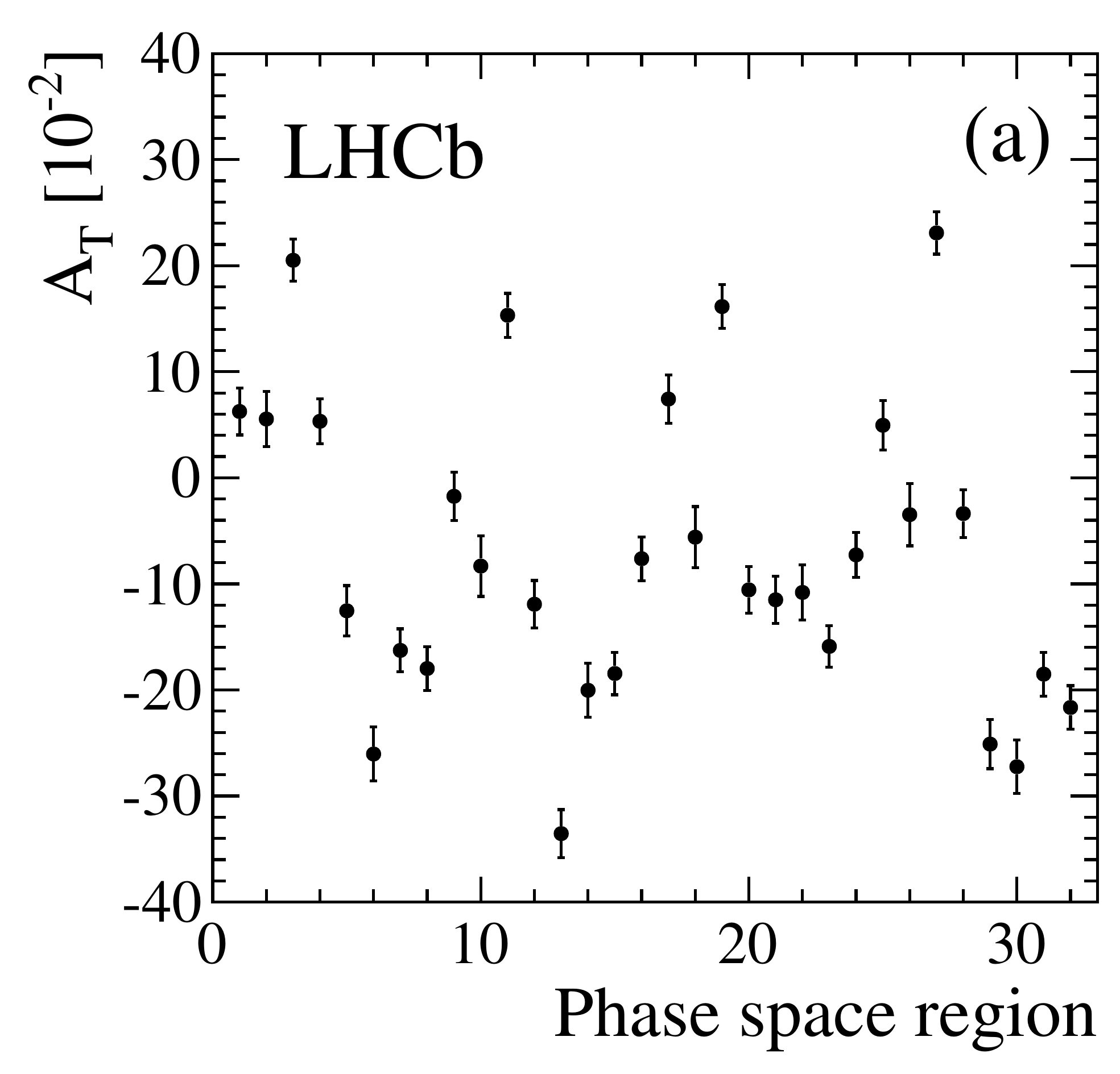}
\includegraphics[width=0.32\textwidth]{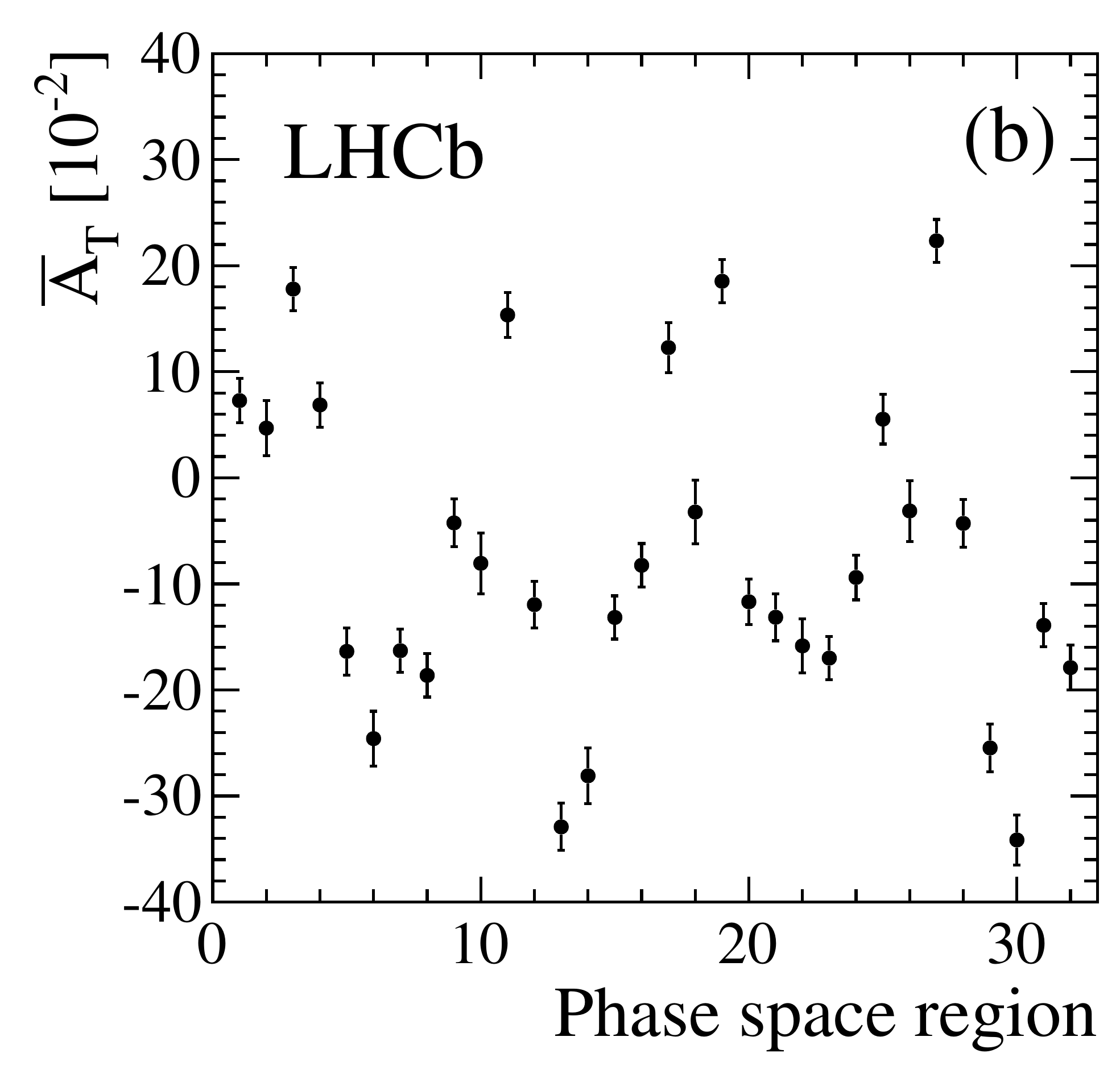}
\includegraphics[width=0.32\textwidth]{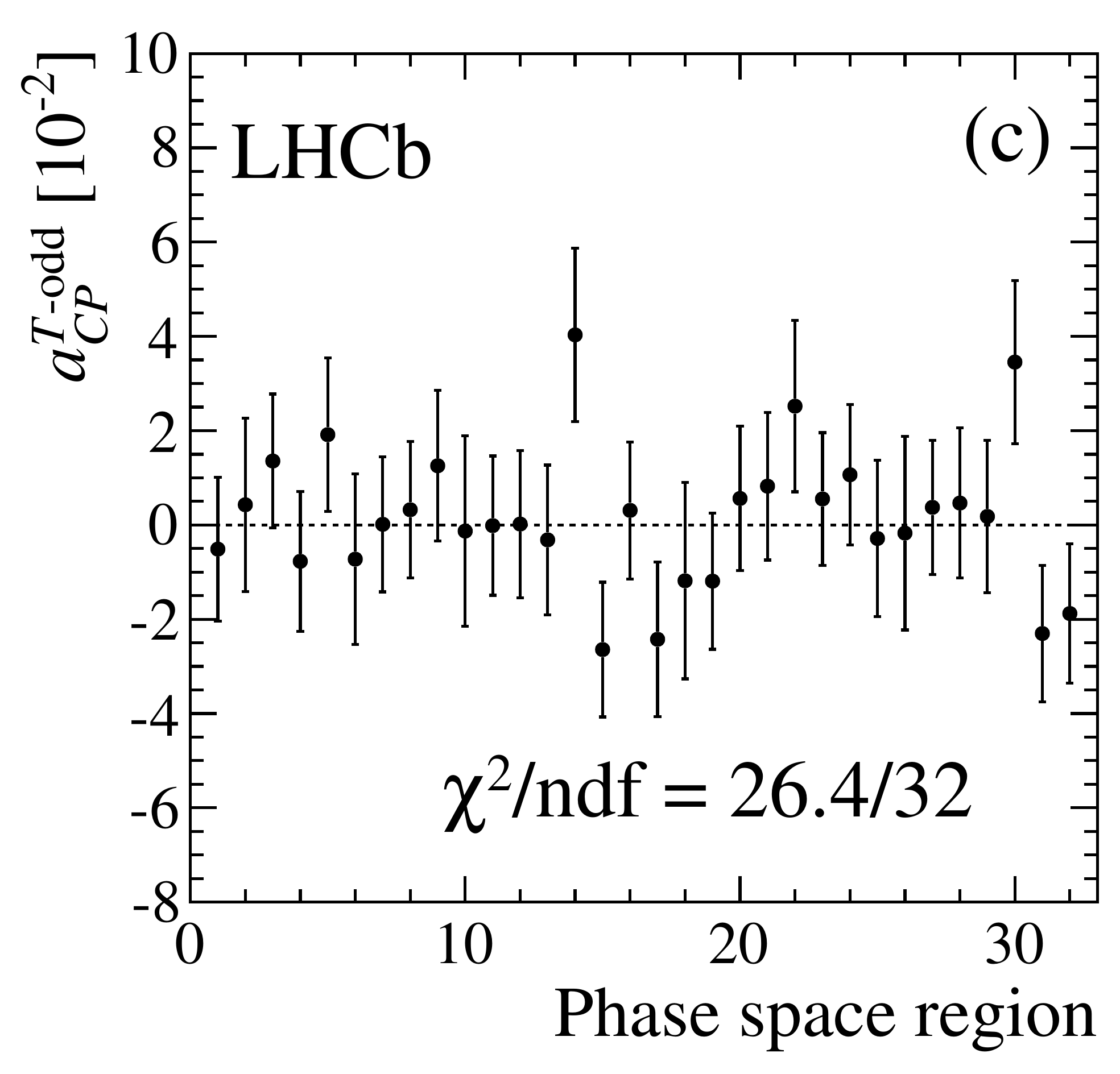}
\caption{\small\label{fig:phspBinning}Distributions of the asymmetry parameters (a) \at , (b) \atb  and (c) \atv  in 32 regions of the phase space.}
\end{figure}

The compatibility with the \CP conservation hypothesis is tested by means of a \chisq test, where the \chisq
 is defined as $R^T V^{-1} R$, where $R$ is the array of \atv measurements, and $V^{-1}$ is the inverse of the covariance matrix $V$, defined as the sum of the statistical and systematic covariance matrices.
An average systematic uncertainty, whose evaluation is discussed in Sec.~\ref{sec:systematics}, is assumed for the different bins.
The statistical uncertainties are considered uncorrelated among the bins, 
while systematic uncertainties are assumed to be fully correlated.
The contribution of systematic uncertainties is small compared to the statistical ones, as shown in Table~\ref{tab:systematics}.
The results are consistent with the \CP conservation hypothesis with a $p$-value of 74\%, 
based on $\chisqndf= 26.4/32$, where ndf is the number of degrees of freedom.
Four alternative binning schemes, one with 8 regions and three with 16 regions, are also tested.
These are described in Appendix~\ref{app:phase_space}.
Results are compatible with the \CP conservation hypothesis with a $p$-value of 24\% for the case of 8 regions 
and 28\%, 62\%, 82\% for the three different phase space divisions in 16 regions.
The \at and \atb asymmetries are significantly different among the different regions.
This effect can be explained by the rich resonant structure of the hadronic 
four-body decay~\cite{Artuso:2012df} that produces different FSI effects over the phase space.

\begin{figure}[tb]
\centering
\small
\includegraphics[width=0.32\textwidth]{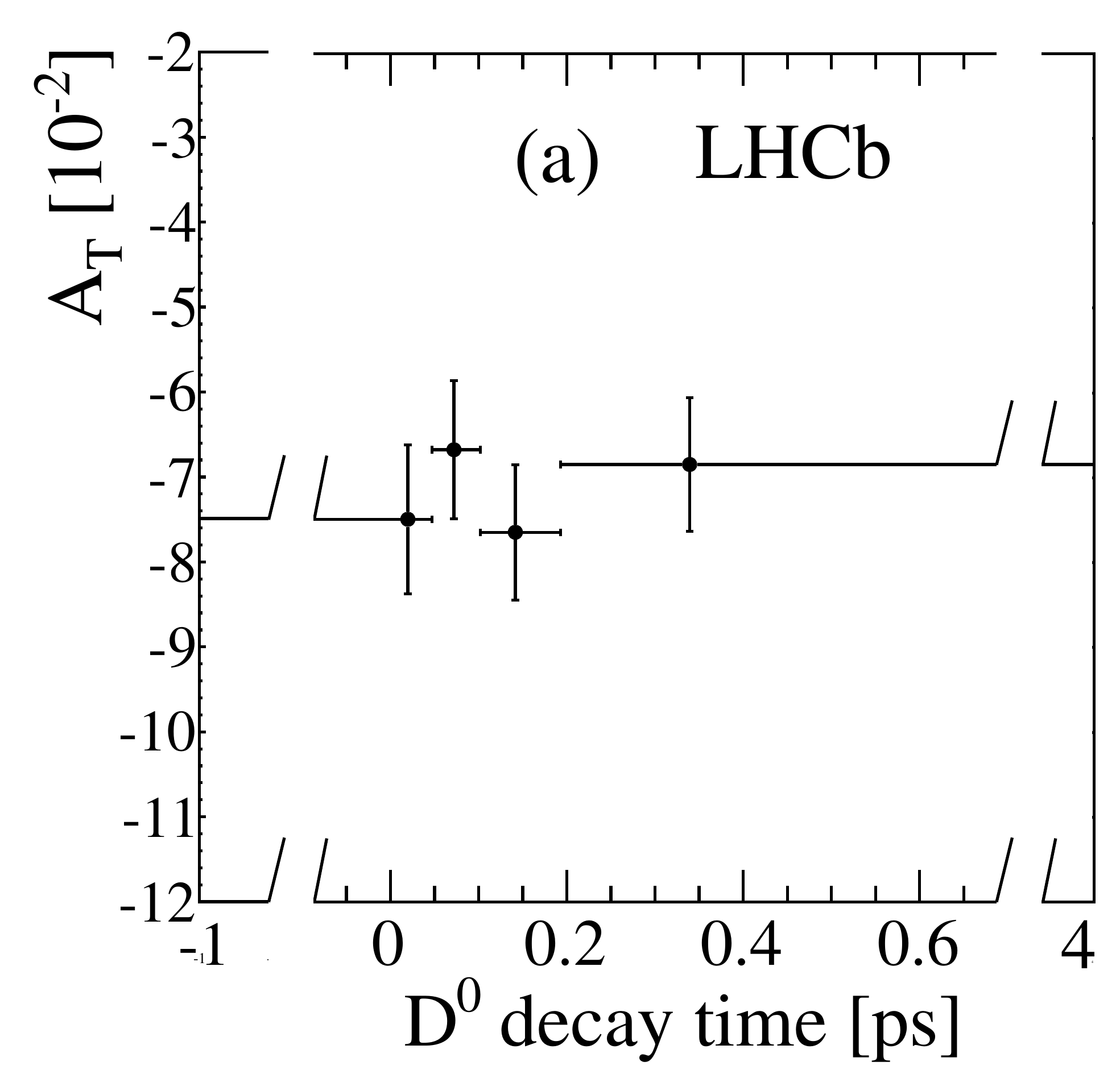}
\includegraphics[width=0.32\textwidth]{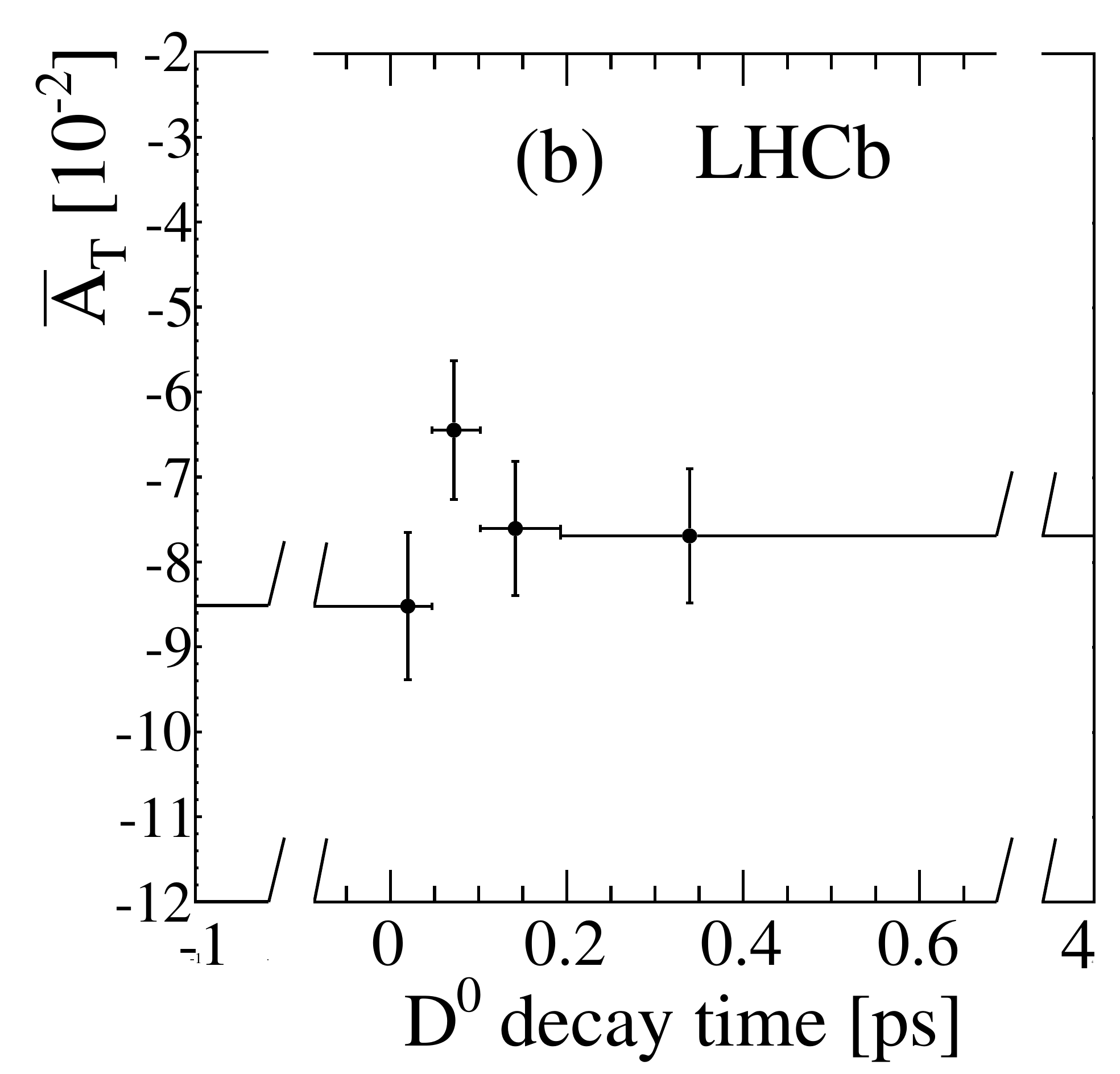}
\includegraphics[width=0.32\textwidth]{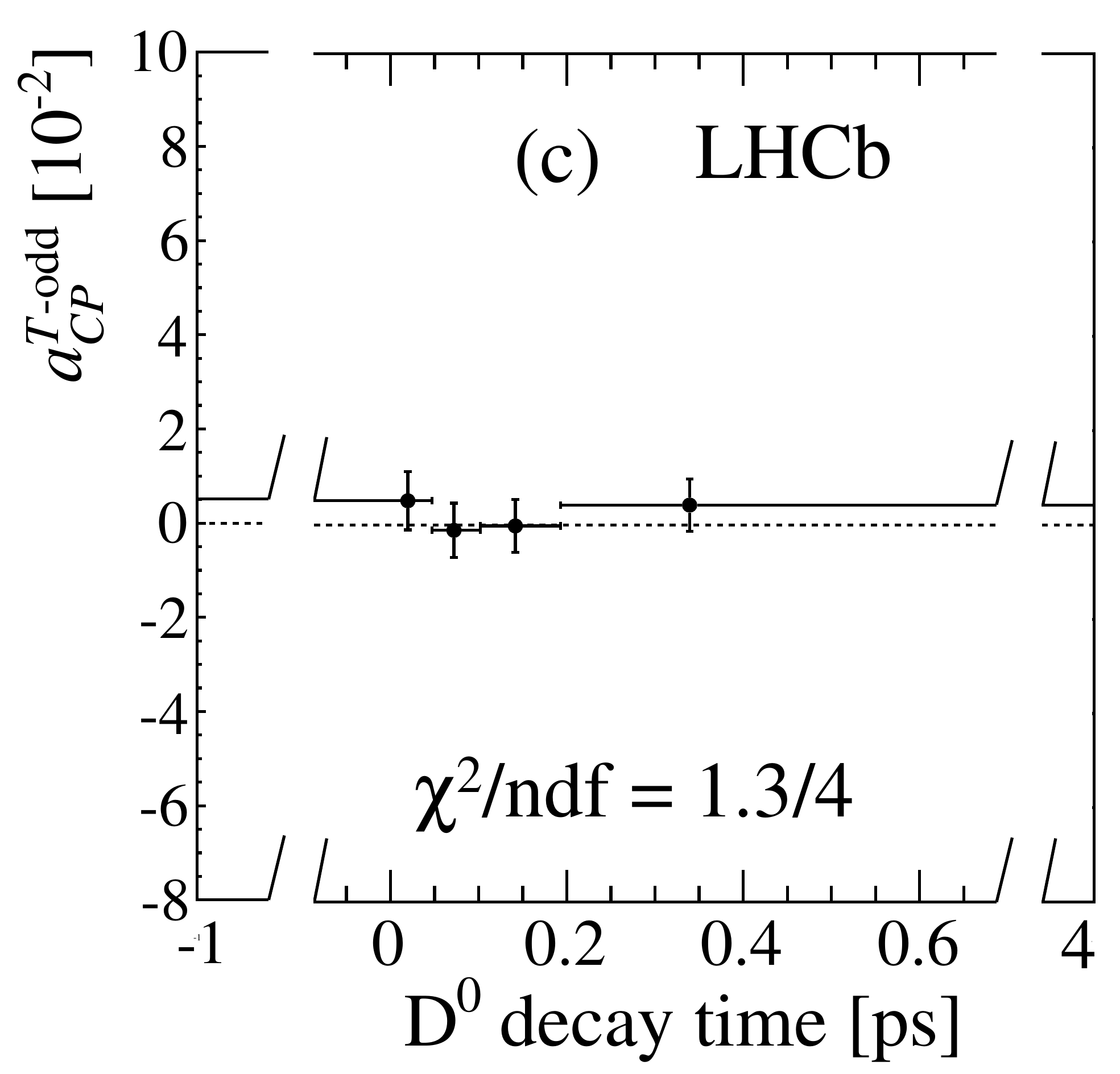}
\caption{\small\label{fig:timeDependent}Distributions of the asymmetry parameters (a) \at, (b) \atb  and (c) \atv  as a function  of the \Dz decay time. For \atv, the value of the \chisqndf for the \CP conservation hypothesis, 
represented by a dashed line, is also quoted.
The scale is broken for the first and last bin.}
\end{figure}

The \atv distribution in \Dz decay time is shown in Fig.~\ref{fig:timeDependent} and the 
results for the different decay time bins are reported in Table~\ref{tab:DecayTime_Results} in Appendix~\ref{app:decay_time}.
The compatibility with the \CP conservation hypothesis is verified by means of a \chisq test considering 
 statistical and systematic uncertainties as in the previous case;
 a value of $\chisqndf= 1.3/4$ is obtained, corresponding to a $p$-value of 86\%. 
Consistent results are obtained when using different divisions of the decay time in 3 and 5 intervals compatible with the \CP conservation hypothesis with $p$-values of 92\% and 83\%, respectively. 
This result is consistent with no time-dependent \CPV.
The \at and \atb asymmetry parameters do not show any significant dependence as a function of the decay time, and the results are compatible with constant functions with $p$-values of 80\% and 38\%, respectively.

\section{Systematic uncertainties}
\label{sec:systematics}

The sources of systematic uncertainty and their relative contributions to the total uncertainty are listed in Table~\ref{tab:systematics}.
\begin{table}[ht]
\small
\centering
\caption{\small\label{tab:systematics}Considered sources of systematic uncertainty and their relative contributions 
to the total uncertainty.}
\begin{tabular}{lccc}
\toprule
Contribution        	& $\Delta \at (\%)$ & $\Delta \atb (\%)$   &  $\Delta \atv (\%)$   \\
\midrule
Prompt background   	& $\pm 0.09$ & $\pm 0.08$   &  $\pm 0.00$   \\
Detector bias   	& $\pm 0.04$ & $\pm 0.04$   &  $\pm 0.04$   \\
\ct resolution      	& $\pm 0.02$ & $\pm 0.03$   &  $\pm 0.01$   \\
Fit model           	& $\pm 0.01$ & $\pm 0.01$   &  $\pm 0.01$   \\
Flavour misidentification& $\pm 0.08$ & $\pm 0.07$   &  $\pm 0.00$   \\
\midrule
Total              		& $\pm 0.13$ & $\pm 0.12$   &  $\pm 0.04$   \\
\bottomrule
\end{tabular}

\end{table}

The contamination from prompt \Dz decays affects the  asymmetry $A$ according to $A \to A(1-f)+f A^d$, where $f$ is the fraction of the contamination in the selected sample and $A^d$ is its own asymmetry. 
The uncertainties are evaluated by using as input the fraction $f_{\text{prompt}}$ 
and the asymmetries of the prompt charm sample. 
These events correspond to random combinations of muons with \Dz originating at the primary vertex 
and show no significant evidence of correlation between the flavour of the \Dz and the charge of the muon.
Assuming that the flavour mistag rate for the prompt \Dz and \Dzb samples is 0.5, 
these are related to the signal asymmetries as: $\at^d = \atv$, $\atb^d=-\atv$, and ${\atv}^d=\atv$.

The detector bias is tested by measuring $\atv(\DzToKpipipi)$ on the $\Dz\to\Km\pip\pip\pim$ control sample.
In this case, a kinematic selection is required between the pair of pions with identical charge to define the \ct and \ctb triple products.
The pion with higher momentum is used to calculate the triple product.
Since this is a Cabibbo-favoured decay, the \CP-violating effects are assumed to be negligible, and any significant deviation from zero is considered as a bias introduced by the experimental technique and the detector reconstruction.
The asymmetry obtained on the control sample is compatible with no bias, $\atv(\DzToKpipipi)=(0.05\pm0.04)\%$. 
A systematic uncertainty equal to the statistical uncertainty of this measurement is assigned.
The test was repeated for different regions of phase space with consistent results.

The systematic uncertainty from detector resolution on \ct is estimated from a simulated sample of $\Dz\to\Kp\Km\pip\pim$ decays where neither FSI nor \CP-violating effects are present.
The difference between the reconstructed and generated asymmetry is considered as a systematic uncertainty due to this effect.

The fit models for signal and background are modified, ensuring good fit quality, 
to account for model-dependent uncertainties.
The signal shape is described with a Gaussian function plus a 
second Gaussian function with a low-mass power-law tail.
The background is described with a third-order polynomial function.
Alternative models are fitted to the data and for each model 1000 simulated samples are generated according to fit results.
The nominal model is then fitted to the simulated samples and the asymmetry parameters are extracted.
Since the bias is not significantly different from zero, its statistical uncertainty is taken as the systematic uncertainty due to this source.

Wrongly identified muon candidates could affect the \CP-violating asymmetry as
\begin{align}
\atv &\to \atv-\Delta\omega/2(\at+\atb),
\end{align}
where $\Delta\omega\equiv\omega^+-\omega^-$ is the difference between the probability of assigning a wrong \Dz ($\omega^+$) 
and \Dzb ($\omega^-$) flavour.
Similar considerations enable the estimation of the uncertainty on \at (\atb) as  
$(\at+\atb)\omega^+$ ($(\at+\atb)\omega^-$).
The mistag probabilities are measured by reconstructing the double-tagged decay channel $\B\to\Dstarp\mu^-X$, with $\Dstarp\to\Dz\pip$ and $\Dz\to\Kp\Km\pip\pim$, and calculating the fraction of events for which the charge of the muon is identical to the charge of the soft pion from \Dstarp decay.
Mistag probabilities of $\omega^{+} = (5.2\pm1.0)\times 10^{-3}$ and $\omega^{-} = (4.7\pm1.0)\times 10^{-3}$ are measured for \Dz and \Dzb flavour, respectively.

Since the various contributions to the systematic uncertainty are independent, the total uncertainty is obtained by summing them in quadrature, and it is very small.
In particular, the \atv observable is insensitive to the production asymmetry of \Dz and \Dzb and to reconstruction-induced charge asymmetries. 
Further cross-checks are made for establishing the stability of the results with respect to the different periods of data taking, different magnet polarities, the choice made in the selection of multiple candidates, and the effect of selection through particle identification criteria.
All these tests reported effects compatible to statistical fluctuations, and therefore are not included in the systematic uncertainty.

The total systematic uncertainties reported in Table~\ref{tab:systematics} are used 
for the measurement of the asymmetries in all regions of phase space 
and in all bins of \Dz decay time.

\section{Conclusions}
\label{sec:conclusions}

In conclusion, a search for \CPV in $\Dz\to\Kp\Km\pip\pim$ decays produced in $\B\to\Dz\mu^-X$ transitions has been performed.
The data sample consists of about $171\,300$ signal decays. 
Three different approaches have been followed to exploit the full potential of the data: 
a measurement integrated over the phase space,  measurements in different regions of the  phase space, 
and measurements as a function of the \Dz decay time.

The results from the phase space integrated measurement,
\begin{align*}
\at&=(-7.18\pm0.41(\text{stat})\pm0.13(\text{syst}))\%,\\
\atb&=(-7.55\pm0.41(\text{stat})\pm0.12(\text{syst}))\%,\\
\atv&=(\phantom{-}0.18\pm0.29(\text{stat})\pm0.04(\text{syst}))\%,
\end{align*}
are consistent with those measured in Ref.~\cite{delAmoSanchez:2010xj}, with significantly improved 
 statistical and systematic uncertainties.
The evaluation of the systematic uncertainties is based mostly on high statistics control samples.

An analysis of the asymmetries in different regions of the phase space is made for the first time and the results are consistent with \CP conservation.
Relatively large variations of \at and \atb over the phase space are measured, which are due to FSI effects produced in the rich resonant structure of the decay~\cite{Donoghue:1986nu,Artuso:2012df}.
For the first time the \atv asymmetry is measured as a function of the \Dz decay time 
and does not show any significant structure at the observed sensitivity.
These results further constrain extensions of the SM~\cite{PhysRevD.75.036008}.

\clearpage

{\noindent\bf\Large Appendices}

\appendix

\section{Measured asymmetries in regions of phase space}
\label{app:phase_space}
The definitions of the 32 regions of phase space of the four-body \DzToKKpipi decay are reported in Table~\ref{tab:32regions}.
The measurements in each region of phase space for the \atv, \at, and \atb  are reported in Table~\ref{tab:PhaseSpace_Results}.%
\begin{table}[ht]
\small
\centering
\caption{\small\label{tab:32regions}Definition of the 32 regions of the 
five-dimensional phase space of the four-body \DzToKKpipi decay.}
\begin{tabular}{lccccc}
\toprule
Region  & $\Phi$          & $m_{\pip\pim}$ (\gevcc)   & $m_{\Kp\Km}$ (\gevcc)  &  $\cos(\theta_{\pip})$  & $\cos(\theta_{\Kp})$    \\
\midrule
1    & (0.00, 1.99)     & (0.20, 0.65)            & (0.60, 1.08)    & ($-$1.00, $-$0.22) & ($-$1.00, $-$0.28) \\
2    & (0.00, 1.99)     & (0.20, 0.65)            & (1.08, 1.70)   & ($-$1.00, $-$0.24) & ($-$1.00, $-$0.14) \\
3    & (0.00, 1.99)     & (0.65, 1.00)            & (0.60, 1.02)   & ($-$1.00, $-$0.09) & ($-$1.00, $-$0.03) \\
4    & (0.00, 1.99)     & (0.65, 1.00)            & (1.02, 1.70)   & ($-$1.00, $-$0.09) & ($-$1.00, $-$0.26) \\
5    & (1.99, 3.14) & (0.20, 0.68)               & (0.60, 1.12)    & ($-$1.00, $-$0.02) & ($-$1.00, $-$0.04) \\
6    & (1.99, 3.14) & (0.20, 0.68)            & (1.12, 1.70)   & ($-$1.00, \phantom{$-$}0.01) & ($-$1.00, \phantom{$-$}0.07) \\
7    & (1.99, 3.14) & (0.68, 1.00)            & (0.60, 1.04)   & ($-$1.00, \phantom{$-$}0.28) & ($-$1.00, $-$0.05) \\
8    & (1.99, 3.14) & (0.68, 1.00)            & (1.04, 1.70)   & ($-$1.00, \phantom{$-$}0.27) & ($-$1.00, $-$0.08) \\
9    & (0.00, 1.99)     & (0.20, 0.65)            & (0.60, 1.08)    & ($-$0.22, \phantom{$-$}1.00) & ($-$1.00, $-$0.28) \\
10    & (0.00, 1.99)     & (0.20, 0.65)            & (1.08, 1.70)   & ($-$0.24, \phantom{$-$}1.00) & ($-$1.00, $-$0.15) \\
11    & (0.00, 1.99)     & (0.65, 1.00)            & (0.60, 1.02)   & ($-$0.09, \phantom{$-$}1.00) & ($-$1.00, $-$0.03) \\
12    & (1.99, 3.14) & (0.20, 0.68)            & (0.60, 1.12)   & ($-$0.02, \phantom{$-$}1.00) & ($-$1.00, $-$0.04) \\
13    & (1.99, 3.14) & (0.20, 0.68)            & (1.12, 1.70)   & (\phantom{$-$}0.01, \phantom{$-$}1.00) & ($-$1.00, \phantom{$-$}0.07) \\
14    & (1.99, 3.14) & (0.68, 1.00)            & (0.60, 1.04)   & (\phantom{$-$}0.28, \phantom{$-$}1.00) & ($-$1.00, $-$0.05) \\
15    & (1.99, 3.14) & (0.68, 1.00)            & (1.04, 1.70)   & (\phantom{$-$}0.27, \phantom{$-$}1.00) & ($-$1.00, $-$0.08) \\
16    & (0.00, 1.99)     & (0.20, 0.65)            & (0.60, 1.08)    & ($-$1.00, \phantom{$-$}0.10) & ($-$0.28, \phantom{$-$}1.00) \\
17    & (0.00, 1.99)     & (0.20, 0.65)            & (1.08, 1.70)   & ($-$1.00, \phantom{$-$}0.01) & ($-$0.15, \phantom{$-$}1.00) \\
18   & (0.00, 1.99)     & (0.65, 1.00)            & (0.60, 1.02)   & ($-$1.00, \phantom{$-$}0.28) & ($-$0.03, \phantom{$-$}1.00) \\
19   & (0.00, 1.99)     & (0.65, 1.00)            & (1.02, 1.70)   & ($-$1.00, $-$0.12) & ($-$0.26, \phantom{$-$}1.00) \\
20   & (1.99, 3.14) & (0.20, 0.68)            & (0.60, 1.12)   & ($-$1.00, \phantom{$-$}0.07) & ($-$0.04, \phantom{$-$}1.00) \\
21   & (1.99, 3.14) & (0.20, 0.68)            & (1.12, 1.70)   & ($-$1.00, \phantom{$-$}0.11) & (\phantom{$-$}0.07, \phantom{$-$}1.00) \\
22   & (1.99, 3.14) & (0.68, 1.00)            & (0.60, 1.04)   & ($-$1.00, $-$0.13) & ($-$0.05, \phantom{$-$}1.00) \\
23   & (1.99, 3.14) & (0.68, 1.00)            & (1.04, 1.70)   & ($-$1.00, $-$0.15) & ($-$0.08, \phantom{$-$}1.00) \\
24    & (0.00, 1.99)     & (0.20, 0.65)            & (0.60, 1.08)    & (\phantom{$-$}0.10, \phantom{$-$}1.00) & ($-$0.28, \phantom{$-$}1.00) \\
25    & (0.00, 1.99)     & (0.20, 0.65)            & (1.08, 1.70)   & (\phantom{$-$}0.01, \phantom{$-$}1.00) & ($-$0.15, \phantom{$-$}1.00) \\
26   & (0.00, 1.99)     & (0.65, 1.00)            & (0.60, 1.02)   & (\phantom{$-$}0.28, \phantom{$-$}1.00) & ($-$0.03, \phantom{$-$}1.00) \\
27   & (0.00, 1.99)     & (0.65, 1.00)            & (1.02, 1.70)   & ($-$0.12, \phantom{$-$}1.00) & ($-$0.26, \phantom{$-$}1.00) \\
28   & (1.99, 3.14) & (0.20, 0.68)            & (0.60, 1.12)   & (\phantom{$-$}0.07, \phantom{$-$}1.00) & ($-$0.04, \phantom{$-$}1.00) \\
29   & (1.99, 3.14) & (0.20, 0.68)            & (1.12, 1.70)   & (\phantom{$-$}0.11, \phantom{$-$}1.00) & (\phantom{$-$}0.07, \phantom{$-$}1.00) \\
30   & (1.99, 3.14) & (0.68, 1.00)            & (0.60, 1.04)   & ($-$0.13, \phantom{$-$}1.00) & ($-$0.05, \phantom{$-$}1.00) \\
31    & (0.00, 1.99)     & (0.65, 1.00)            & (1.02, 1.70)   & ($-$0.10, \phantom{$-$}1.00) & ($-$1.00, $-$0.26) \\
32   & (1.99, 3.14) & (0.68, 1.00)            & (1.04, 1.70)   & ($-$0.15, \phantom{$-$}1.00) & ($-$0.08, \phantom{$-$}1.00) \\
\bottomrule
\end{tabular}

\end{table}
\begin{table}[ht]
\small
\centering
\caption{\small \label{tab:PhaseSpace_Results}Measurements of \atv, \at and \atb in each region 
of phase space. The uncertainties are statistical only. A common systematic uncertainty of 0.13\%, 0.12\% and 0.04\% should be added to the asymmetries \at, \atb and \atv, respectively. This uncertainty is considered fully correlated among the bins.}
\begin{tabular}{lccccccc}
\toprule
     &  Region 1          & Region 2            & Region 3          & Region 4            & Region 5\\
\midrule
\atv (\%) & $-0.51\pm1.53$ & $\phantom{-}0.43\pm1.84$  & $\phantom{2}1.36\pm1.42$ & $-0.77\pm1.48$   & $\phantom{-1}1.91\pm1.62$   \\
\at \phantom{.11}(\%) & $\phantom{-}6.25\pm2.21$  & $\phantom{-}5.54\pm2.59$ & $20.51\pm1.98$ & $\phantom{-}5.33\pm2.10$    & $-12.54\pm2.37$  \\
\atb \phantom{.11}(\%) & $\phantom{-}7.28\pm2.10$  & $\phantom{-}4.68\pm2.61$ & $17.79\pm2.04$ & $\phantom{-}6.87\pm2.09$    & $-16.37\pm2.22$  \\
\midrule
       & Region 6   & Region 7           & Region 8  &   Region 9          & Region 10\\            
\midrule
\atv (\%)  & $\phantom{}-0.73\pm1.81$& $\phantom{-1}0.01\pm1.43$  & $\phantom{-1}0.32\pm1.45$ & $\phantom{-}1.26\pm1.60$ & $-0.13\pm2.02$  \\
\at \phantom{.11}(\%)   & $-26.04\pm2.54$ 	& $-16.27\pm2.03$ & $-17.99\pm2.06$  & $-1.74\pm2.27$  & $-8.32\pm2.86$ \\
\atb \phantom{.11}(\%)  & $-24.59\pm2.58$	& $-16.30\pm2.03$ & $-18.63\pm2.04$  &  $-4.25\pm2.25$  & $-8.06\pm2.86$  \\
\midrule
       & Region 11          & Region 12  	& Region 13          & Region 14          & Region 15\\
\midrule
\atv (\%)  & $-0.01\pm1.48$ & $\phantom{-1}0.02\pm1.56$   & $\phantom{}-0.32\pm1.59$   & $\phantom{-1}4.03\pm1.84$   & $\phantom{}-2.64\pm1.43$    \\
\at \phantom{.11}(\%)   & $\phantom{.}15.32\pm2.07$ & $-11.92\pm2.24$ 	& $-33.55\pm2.26$ & $-20.03\pm2.55$ & $-18.45\pm2.00$  \\
\atb \phantom{.11}(\%)  & $\phantom{.}15.35\pm2.11$ & $-11.96\pm2.18$	& $-32.91\pm2.22$ & $-28.09\pm2.64$ & $-13.17\pm2.04$   \\
\midrule
       & Region 16    & Region 17      & Region 18 & Region 19          & Region 20 \\
\midrule
\atv (\%)  & $\phantom{-}0.31\pm1.45$ & $-2.43\pm1.64$ & $-1.18\pm2.08$  & $-1.19\pm1.45$ & $\phantom{-1}0.56\pm1.53$   \\
\at \phantom{.11}(\%)  & $-7.63\pm2.06$    & $\phantom{-}7.42\pm2.28$  & $-5.60\pm2.89$& $\phantom{.}16.15\pm2.05$ & $-10.57\pm2.19$  \\
\atb \phantom{.11}(\%) & $-8.25\pm2.05$    & $\phantom{.}12.28\pm2.35$  & $-3.23\pm3.00$ & $\phantom{.}18.53\pm2.04$ & $-11.69\pm2.15$  \\
\midrule
      & Region 21           & Region 22    & Region 23      & Region 24 &  Region 25    \\
\midrule
\atv (\%) & $\phantom{-1}0.82\pm1.57$   & $\phantom{-1}2.52\pm1.82$   & $\phantom{-1}0.55\pm1.41$  & $\phantom{-}1.06\pm1.49$ & $-0.29\pm1.66$ \\
\at  \phantom{.11}(\%) & $-11.50\pm2.22$ & $-10.81\pm2.59$ & $-15.90\pm1.96$ & $-7.27\pm2.11$ & $\phantom{-}4.95\pm2.34$\\
\atb \phantom{.11}(\%) & $-13.15\pm2.21$ & $-15.85\pm2.55$ & $-17.00\pm2.02$ & $-9.40\pm2.11$ & $\phantom{-}5.53\pm2.34$\\
\midrule
      & Region 26            & Region 27          & Region 28            & Region 29          & Region 30\\
\midrule
\atv (\%) & $-0.17\pm2.05$  & $\phantom{1}0.37\pm1.42$ & $\phantom{-}0.46\pm1.59$   & $\phantom{-1}0.18\pm1.62$   & $\phantom{-1}3.45\pm1.73$ \\
\at \phantom{.11}(\%)  & $-3.47\pm2.94$ & $23.09\pm2.01$ & $-3.37\pm2.26$ & $-25.11\pm2.32$ & $-27.24\pm2.53$ \\
\atb \phantom{.11}(\%) & $-3.13\pm2.87$ & $22.34\pm2.02$ & $-4.30\pm2.25$ & $-25.47\pm2.25$ & $-34.15\pm2.36$ \\
\midrule
      & Region 31           & Region 32    & & &  \\
\midrule
\atv (\%) & $\phantom{}-2.30\pm1.45$  & $\phantom{}-1.88\pm1.47$ & & & \\
\at  \phantom{.11}(\%) & $-18.51\pm2.06$ & $-21.65\pm2.05$ & & & \\
\atb \phantom{.11}(\%) & $-13.91\pm2.04$ & $-17.89\pm2.12$ & & &      \\
\bottomrule
\end{tabular}

\end{table}

The alternative binning schemes with 8 and 16 bins have been defined by integrating over $\cos(\theta_{\Kp})$ and  $\cos(\theta_{\pip})$ (8 bins), $\cos(\theta_{\Kp})$ (16 bins), $\cos(\theta_{\pip})$ (16 bins), and by using mass variables ($m_{\Km\pip}$, $m_{\Kp\Km\pip}$, $m_{\Km\pip\pim}$) in place of the angular variables (16 bins).

\section{Measured asymmetries in intervals of \Dz decay time}
\label{app:decay_time}
The values measured in 4 different intervals of the \Dz decay time for \atv, \at, and \atb 
 are reported in Table~\ref{tab:DecayTime_Results}.
\begin{table}[ht]
\small
\centering
\caption{\small \label{tab:DecayTime_Results}Measurements of \atv, \at and \atb in different intervals of \Dz decay time, $t$, expressed in \ps. 
The uncertainties are statistical only. 
A common systematic uncertainty of 0.13\%, 0.12\% and 0.04\% should be added to the asymmetries \at, \atb and \atv, respectively. This uncertainty is considered fully correlated among the bins.}
\begin{tabular}{lcccc}
\midrule
     &  $[-1.00,0.05]$  & $[0.05,0.10]$   &  $[0.10,0.19]$ &  $[0.19,4.00]$  \\
\midrule
\atv (\%) & $\phantom{-}0.51\pm0.62$   & $-0.12\pm0.58$  & $-0.02\pm0.56$ & $\phantom{-}0.42\pm0.56$  \\
\at \phantom{.11}(\%) & $-7.49\pm0.88$  & $-6.67\pm0.81$ & $-7.64\pm0.80$  &  $-6.84\pm0.79$  \\
\atb \phantom{.11}(\%) & $-8.51\pm0.87$  & $-6.44\pm0.82$ & $-7.60\pm0.79$  & $-7.68\pm0.79$  \\
\bottomrule
\end{tabular}

\end{table}

\section*{Acknowledgements}

\noindent We express our gratitude to our colleagues in the CERN
accelerator departments for the excellent performance of the LHC. We
thank the technical and administrative staff at the LHCb
institutes. We acknowledge support from CERN and from the national
agencies: CAPES, CNPq, FAPERJ and FINEP (Brazil); NSFC (China);
CNRS/IN2P3 (France); BMBF, DFG, HGF and MPG (Germany); SFI (Ireland); INFN (Italy); 
FOM and NWO (The Netherlands); MNiSW and NCN (Poland); MEN/IFA (Romania); 
MinES and FANO (Russia); MinECo (Spain); SNSF and SER (Switzerland); 
NASU (Ukraine); STFC (United Kingdom); NSF (USA).
The Tier1 computing centres are supported by IN2P3 (France), KIT and BMBF 
(Germany), INFN (Italy), NWO and SURF (The Netherlands), PIC (Spain), GridPP 
(United Kingdom).
We are indebted to the communities behind the multiple open 
source software packages on which we depend. We are also thankful for the 
computing resources and the access to software R\&D tools provided by Yandex LLC (Russia).
Individual groups or members have received support from 
EPLANET, Marie Sk\l{}odowska-Curie Actions and ERC (European Union), 
Conseil g\'{e}n\'{e}ral de Haute-Savoie, Labex ENIGMASS and OCEVU, 
R\'{e}gion Auvergne (France), RFBR (Russia), XuntaGal and GENCAT (Spain), Royal Society and Royal
Commission for the Exhibition of 1851 (United Kingdom).

\clearpage

\addcontentsline{toc}{section}{References}
\setboolean{inbibliography}{true}
\bibliographystyle{LHCb}
\bibliography{main,LHCb-PAPER,LHCb-CONF,LHCb-DP,LHCb-TDR}

\newpage
\centerline{\large\bf LHCb collaboration}
\begin{flushleft}
\small
R.~Aaij$^{41}$, 
B.~Adeva$^{37}$, 
M.~Adinolfi$^{46}$, 
A.~Affolder$^{52}$, 
Z.~Ajaltouni$^{5}$, 
S.~Akar$^{6}$, 
J.~Albrecht$^{9}$, 
F.~Alessio$^{38}$, 
M.~Alexander$^{51}$, 
S.~Ali$^{41}$
% ...

{\footnotesize \it
% ...
$ ^{5}$Clermont Universit\'{e}, Universit\'{e} Blaise Pascal, CNRS/IN2P3, LPC, Clermont-Ferrand, France\\
$ ^{6}$CPPM, Aix-Marseille Universit\'{e}, CNRS/IN2P3, Marseille, France\\
% ...
$ ^{9}$Fakult\"{a}t Physik, Technische Universit\"{a}t Dortmund, Dortmund, Germany\\
% ...
$ ^{37}$Universidad de Santiago de Compostela, Santiago de Compostela, Spain\\
$ ^{38}$European Organization for Nuclear Research (CERN), Geneva, Switzerland\\
% ...
$ ^{41}$Nikhef National Institute for Subatomic Physics, Amsterdam, The Netherlands\\
% ...
$ ^{46}$H.H. Wills Physics Laboratory, University of Bristol, Bristol, United Kingdom\\
% ...
$ ^{51}$School of Physics and Astronomy, University of Glasgow, Glasgow, United Kingdom\\
$ ^{52}$Oliver Lodge Laboratory, University of Liverpool, Liverpool, United Kingdom\\
% ...
}
\end{flushleft}

\end{document}